\documentclass[aps,
prd,
showpacs,
showkeys,
amsmath,
amssymb,
twocolumn,
floatfix,
superscriptaddress
]{revtex4-1}
\usepackage{graphicx}
\usepackage{times}
\usepackage{verbatim}
\usepackage{color}
\usepackage{cancel}
\usepackage[normalem]{ulem}
\usepackage{multirow}
\usepackage{braket}
\usepackage{bm}
\usepackage{hyperref}
\usepackage{tabularx}
\usepackage[capitalise]{cleveref}
\usepackage[subpreambles=true,sort]{standalone}
\usepackage{slashed}

\newcommand{\Dubna}{\affiliation{Joint~Institute~for~Nuclear~Research, Dubna, Moscow~Region}}

\newcommand{\normord}[1]{%
	:\mathrel{\mspace{1mu}#1\mspace{1mu}}:%
}

\begin{document}

\title{Coherency and incoherency in neutrino-nucleus elastic and inelastic scattering}

\author{Vadim~A.~Bednyakov}\Dubna
\author{Dmitry~V.~Naumov}\Dubna
\date{\today}

\begin{abstract}
\noindent
Neutrino-nucleus scattering $\nu A\to \nu A$, in which the nucleus conserves its integrity, is considered.
Our consideration follows a microscopic description of the nucleus as a bound state of its constituent nucleons described by a multi-particle wave-function of a general form.

We  show  that elastic interactions keeping the nucleus in the same quantum state lead to a quadratic enhancement of the corresponding cross-section in terms of the number of nucleons.
Meanwhile, the cross-section of inelastic processes in which the quantum state of the nucleus is changed, essentially has a linear dependence on the number of nucleons.
These two classes of processes are referred to as coherent and incoherent, respectively.

Accounting for all possible  initial and final internal states of the nucleus leads to a general conclusion independent of the nuclear model.
The coherent and incoherent cross-sections are driven by factors $|F_{p/n}|^2$ and $(1-|F_{p/n}|^2)$, where $|F_{p/n}|^2$ is a proton/neutron form-factor of the nucleus, averaged over its initial states.
Therefore, our assessment suggests a smooth transition between regimes of coherent and incoherent neutrino-nucleus scattering.
In general, both  regimes contribute to experimental observables.

The coherent cross-section formula used in the literature is revised and corrections depending on kinematics are estimated.
Consideration of only those matrix elements  which correspond to the same initial and final  spin states of the nucleus and accounting for a non-zero momentum of the target nucleon are two main sources of the corrections.

As an illustration of the importance of the incoherent channel we considered three experimental setups with different nuclei.
As an  example, for ${}^{133}\text{Cs}$ and neutrino energies of $30-50$ MeV the incoherent cross-section is about 10-20\% of the coherent contribution if experimental detection threshold is accounted for.

Experiments attempting to measure coherent neutrino scattering by solely detecting the recoiling nucleus, as is typical, might be including an incoherent background that is indistinguishable from the signal if the excitation gamma eludes its detection. 
However, as is shown here, the incoherent component can be measured directly by searching for photons released by the excited nuclei inherent to the incoherent channel.
For a beam experiment these gammas should be correlated in time with the beam, and their higher energies make the corresponding signal easily detectable at a rate governed by the ratio of incoherent to coherent cross-sections.
The detection of signals due to the nuclear recoil and excitation $\gamma$s provides a more sensitive instrument in studies of nuclear structure and possible signs of new physics.
\end{abstract}

\maketitle
\tableofcontents
\section{Introduction}
The process of neutrino scattering, by means of $Z^0$-boson exchange, off a  system of bonded particles provides a great laboratory to test principles of quantum physics and search for new phenomena.
Under certain conditions the corresponding interaction probability acquires an extra factor with respect to the case of scattering off free particles.

This extra factor, proportional to the number of scatterers, is a direct consequence of the principles of quantum physics.
The probability of an outcome is determined by the absolute value squared of the sum of amplitudes corresponding to indistinguishable paths to realize this outcome.
Neutrino-nucleus scattering in which the nucleus conserves its integrity  is an example of this kind, as was observed by Freedman~\cite{Freedman:1973yd} more than four decades ago.

There are two distinct outcomes of such interactions: (i) the nucleus remains in the same quantum state and (ii) the state is changed.
We refer to these cases as elastic and inelastic scatterings, respectively, because in (i) the energy transfer to the recoil nucleus is vanishingly small, while in (ii) it is apparently non-zero.

It was shown~\cite{Drukier:1983gj,Barranco:2005yy,Patton:2012jr,Papoulias:2015vxa} that the cross-section of elastic neutrino scattering off a nucleus is amplified with respect to a neutrino scattering off a single nucleon.
The amplification factor for  a spin-less even-even nucleus reads
\begin{equation}
\label{eq:enhancement_factor_naive}
\left|g_V^nN F_n(\bm{q})+g_V^pZF_p(\bm{q})\right|^2\simeq N^2 (g_V^n)^2|F_n(\bm{q})|^2,
\end{equation}
where $Z$ and $N$ are the numbers of protons and neutrons, $g_V^{p/n}$ are proton/neutron couplings of the nucleon vector current, and $F_{p/n}(\bm{q})$ are proton/neutron form-factors of the nucleus.
The form-factors approach unity if 
\begin{equation}
\label{eq:qR_condition}
|\bm{q}|R_A\ll 1,
\end{equation}
where $R_A$ is the radius of the nucleus.
The form-factors vanish at $|\bm{q}|\to\infty$.

Neutrinos with energies below some tens of MeV predominately  conserve the integrity of nucleons in neutrino-quark interactions with $Z^0$-boson exchange, allowing one to consider this process using an effective neutrino-nucleon interaction in which the nucleon current is a sum of vector and axial currents.

The corresponding axial currents do not contribute significantly when a neutrino elastic scatters off of a spin-less nucleus due to the cancellation  in the sum of amplitudes.
The vector coupling $g_V^p=\frac{1}{2}-2\sin^2\theta_W$ of the proton is small ($g_V^p\approx 0.023$) and is neglected in the approximate equality in~\cref{eq:enhancement_factor_naive}.
In our estimates we used a best-fit value of $\sin^2\theta_W=0.23865$, determined using low energy neutrino data and $\overline{\text{MS}}$ renormalization scheme~\cite{Kumar:2013yoa}.

Freedman coined the terminology ''coherent neutrino-nucleus scattering'' to emphasize the fact that the dependence of the corresponding cross-section is quadratic in terms of the number of nucleons.
This dependence was attributed to nearly identical amplitude phases corresponding to a neutrino scattering off nucleons.

His first  calculations~\cite{Freedman:1973yd} were revised in a number of papers~\cite{Drukier:1983gj,Barranco:2005yy,Patton:2012jr,Papoulias:2015vxa,Smith:1985mta,Jachowicz:2001jr,Divari:2010zz,McLaughlin:2015xfa}.
The impact of  the nuclear structure models was studied in~\cite{Engel:1991wq,Amanik:2009zz,Amanik:2007ce,Lykasov:2007iy,Patton:2013nwa}.
The importance of the coherent cross-section was examined theoretically for a number of observables in astrophysics, like stellar collapse~\cite{Wilson:1974zz,Freedman:1977xn}, Supernovae~\cite{Bernabeu:1975tw,Rombouts:1997vm,Divari:2012zz,Divari:2012cj} and in studies of physics beyond the Standard Model (SM)~\cite{Barranco:2005yy,Scholberg:2005qs,deNiverville:2015mwa,Esteban:2018ppq,Abdullah:2018ykz,Farzan:2018gtr,Billard:2018jnl,Denton:2018xmq,Ge:2017mcq,Kosmas:2017tsq,Canas:2018rng,AristizabalSierra:2018eqm}, electromagnetic properties of the neutrino~\cite{Papavassiliou:2005cs}, searches for sterile neutrinos~\cite{Formaggio:2011jt,Anderson:2012pn}, and estimates of neutron density in the nucleus~\cite{Patton:2012jr,Cadeddu:2017etk}. 
Coherent scattering of atomic systems was studied in~\cite{Gaponov:1977gr,Sehgal:1986gn}, where the concept of neutrino optics was suggested for neutrinos with energies $\lesssim 10$ keV.

Dating back to the seminal paper by Freedman~\cite{Freedman:1973yd}, a number of  experimental proposals~\cite{Lewis:1979mu,Drukier:1983gj,Horowitz:2003cz,Giomataris:2005fx,Wong:2005vg,Vergados:2009ei,Sangiorgio:2012vsa,Brice:2013fwa,Kopylov:2013zda,Kopylov:2014xra,Agnolet:2016zir,Aguilar-Arevalo:2016khx,Moroni:2014wia,Belov:2015ufh,Tayloe:2017edz,Billard:2016giu} using reactor and accelerator neutrinos were suggested to observe neutrino-nucleus coherent scattering.
This process is an unavoidable background in sensitive searches for dark matter~\cite{Wong:2010zzc,Anderson:2011bi,Gutlein:2014gma,Bednyakov:2015uoa,Fallows:2018ika}.
The difficulty in observing coherent neutrino scattering lies in the detection of scattered nuclei with low kinetic energy of the order of some keV or tens of keV.

The first experimental evidence for coherent neutrino-nucleus scattering was reported in 2017 by the COHERENT Collaboration~\cite{Bolozdynya:2012xv,Akimov:2015nza,Collar:2014lya}, using  CsI[Na] scintillator exposed to  neutrinos with energies of tens of MeV produced by the Spallation Neutron Source (SNS) at the Oak Ridge National Laboratory~\cite{Akimov:2017ade,Akimov:2018vzs,Akimov:2018ghi}.

Our motivation for this work was triggered by the following observation.
At neutrino energies of some tens of MeV the three-momentum transfer $\bm{q}$ is large enough to break the condition in~\cref{eq:qR_condition}.
For example,   energy deposits observed in~\cite{Akimov:2017ade}, correspond to $|\bm{q}|R_A$ sampling the interval $(1,2.7)$ and the elastic cross-section should be suppressed.
At higher energies, but still in the regime where the nucleus conserves its integrity, the elastic cross-section vanishes and the neutrino-nucleus interaction probability must be determined by inelastic interactions. 
In general, the corresponding cross-section  should be given by a sum of elastic and inelastic cross-sections, similar to the theory of the scattering of $X$-rays~\cite{Waller119} and electrons~\cite{Morse1932} off an atom, and of slow neutrons off of matter constituents~\cite{PhysRev.95.249}.

What  should one expect about the ''coherency'' in inelastic processes?
If this terminology is understood literally as the equality of phases of neutrino-nucleon scattering amplitudes, then one would conclude that inelastic processes should also be coherent, as in elastic processes, because there is no reason why these phases should be different.
Should one then expect a quadratic dependence of the inelastic cross-section in terms of the number of nucleons, similar to~\cref{eq:enhancement_factor_naive}?
The corresponding literature, to best of our knowledge, lacks an appropriate theory for neutrino-nucleus interactions that could address these questions.
This paper attempts to provide a theoretical framework accounting for elastic and inelastic neutrino-nucleus  scattering of the process
\begin{equation}
\label{eq:neutrino_nucleus_scattering_definition}
\nu A \to \nu A^{(*)},
\end{equation}
based on calculations from first principles. 
In~\cref{eq:neutrino_nucleus_scattering_definition} the possibility that the  internal quantum state of a nucleus can be modified after an interaction is labeled by the $(*)$ superscript.

We show in this work that the cross-section of the neutrino-nucleus elastic process is, indeed,  quadratically dependent on the number of nucleons, while that for inelastic scattering exhibits a linear dependence.
Elastic and inelastic cross-sections also possess a distinct dependence on $\bm{q}$: the former is driven by $|F_{p/n}|^2$, while the latter is governed by $1-|F_{p/n}|^2$.
At the same time, the phases of corresponding neutrino-proton and neutrino-neutron amplitudes are all equal for protons and neutrons, respectively.
This is at odds with the assumption that the difference of phases of the scattering amplitudes is responsible for loss of coherency~\cite{Freedman:1977xn,Papavassiliou:2005cs,Kerman:2016jqp}.
Our arguments are discussed in what follows.
 
The paper is split into two parts. 
The first part is focused only on the main points of the derivation and discusses the results obtained. 
The second part, containing the necessary technical details, is organized in a set of appendices. 

In particular, a conceptual derivation of a general form of the cross-section of the process in~\cref{eq:neutrino_nucleus_scattering_definition} is discussed  in~\cref{sec:cross-section}.
We review the paradigm of coherent scattering and suggest our concept in a simplified way in~\cref{sec:paradigm}.
The kinematics of elastic and inelastic scattering, and the corresponding amplitude and the cross-section are discussed in~\cref{sec:kinematics,sec:kinematic_paradox,sec:amplitude,sec:cross_section}, respectively.
We refer to~\cref{app:framework,app:cross_section,app:matrix_element_calculation} for full details of this derivation. 

In~\cref{app:framework} we define the theoretical framework,  reminding the reader of the decomposition of a quantum state in $x$ and $p$ bases for $n-$particle states,  introducing notation and defining a general form of the wave-function of the nucleus.
In~\cref{app:cross_section} we compute the scattering amplitude and the cross-section.
In~\cref{app:matrix_element_calculation} we summarize some details of our calculations of the scalar product of lepton and hadron currents, needed to calculate the scattering amplitude and the cross-section.  

In~\cref{sec:discussion} we discuss in detail the derived cross-section.
Coherent and incoherent regimes are discussed in~\cref{sec:coherent_incoherent}. 
Our revision of the coherent cross-section is discussed in~\cref{sec:spin_axial}.
In~\cref{sec:gamma_proposal} we discuss in some detail a proposal to detect transition $\gamma$s from excited nuclei inherent to incoherent processes. 
These $\gamma$s would provide both an additional background suppression and an independent observable sensitive to the form-factor of the nucleus.
In~\cref{app:mechanics} we provide an  analogy with a mechanical system of two balls connected by a spring  to illustrate the kinematics of coherent and incoherent scattering.
The summary is drawn in~\cref{sec:summary}.

The natural units $\hbar=c=1$ are used throughout the paper. 
Three-vectors are denoted by bold face. 
A four-vector  $a$ has the following components: $a^\mu = (a^0,\bm{a})$, enumerated by a Greek index $\mu$. 
The Dirac spinors and $\gamma$-matrices are used in the Dirac basis and $\gamma_5 = i\gamma_0\gamma_1\gamma_2\gamma_3$. 
The Feynman slash notation $\slashed{a} = \gamma^\mu a_\mu$ is used for a scalar product of a four-vector $a_\mu$ and Dirac $\gamma^\mu$-matrices.
Quantum operators are denoted by the hat symbol, like $\hat{\bm{X}}$ for the position operator.

\section{Elastic and inelastic neutrino-nucleus scattering}
\label{sec:cross-section}	
\subsection{Revising the paradigm}
\label{sec:paradigm}
We begin this section by reminding the reader of the paradigm of coherency  in neutrino-nucleus scattering~\cite{Freedman:1977xn}. 
Two waves are considered coherent if they have the same frequencies, wave-forms, and constant relative phase.
Coherence can lead to constructive and destructive interference.

A neutrino-nucleus interaction is a result of an individual neutrino scattering off of nucleons.
Each such scattering off a $k$-th nucleon can be described by an amplitude $\mathcal{A}^k$.

If these nucleons are assumed to have definite coordinates $\bm{x}_k$, then, due to the translation invariance, $\mathcal{A}^k$ gets an additional factor $e^{i\bm{q}\bm{x}_k}$ and the total amplitude reads
\begin{equation}
\mathcal{A} = \sum_{k=1}^A \mathcal{A}^k e^{i\bm{q}\bm{x}_k}.
\label{eq:amplitude_plane_wave}
\end{equation}
These individual amplitudes are coherent if for any $k$ the phases $\bm{q}\bm{x}_k$ are nearly the same.
This is fulfilled if the condition in~\cref{eq:qR_condition} is satisfied.

\begin{figure}[!h]
	\resizebox{0.4\linewidth}{!}{\tikzset{
  particlepath0/.style = {dotted,decoration={markings,mark=at position 0.50 with {\arrow[xshift=0.8mm]{stealth}}},postaction={decorate}},
  particlepath1/.style = {decoration={markings,mark=at position 0.50 with {\arrow[xshift=0.8mm]{stealth}}},postaction={decorate}},
  front0/.style = {},
  front1/.style = {thin,dashed,opacity=0.6},
  arc/.style = {thin,opacity=0.6},
  textstyle/.style = {scale=1.8}
}
\begin{tikzpicture}[very thick]
  \coordinate (start) at (0,0);

  \def\Length{3cm}
  \def\Angle{45}
  \def\MarkerSize{1pt}
  \def\ArcRad{5mm}

  \draw[front0]        (start)  ++(0:\Length)    -- ++(-90:3*\Length);
  \draw[particlepath0] (start) ++(-90:\Length)   -- ++(0:\Length) coordinate (nu1) node[textstyle,midway,above] {$\nu$};
  \draw[particlepath0] (start) ++(-90:2*\Length) -- ++(0:\Length) coordinate (nu2) node[textstyle,midway,below] {$\nu$};

  \draw[fill] (nu1) circle (\MarkerSize) node[textstyle,below left,xshift=1mm] {$\bm{x}_j$};
  \draw[front1] (nu1) -- ++(0:2.5*\ArcRad);
  \draw[fill] (nu2) circle (\MarkerSize) node[textstyle,below left,xshift=1mm] {$\bm{x}_k$};

  \draw[particlepath1] (nu1) -- ++(\Angle:\Length) node[textstyle,right] {$\nu$};
  \draw[particlepath1] (nu2) -- ++(\Angle:\Length) node[textstyle,right] {$\nu$};

  \draw[stealth-stealth,thin] (nu2) ++(-\Angle:3mm) -- ++(\Angle:\Length/1.41421356) node [midway,rotate=\Angle,below] {$\Delta \varphi$};

  \draw[front1] (nu1) ++(\Angle+90:\Length) -- ++(\Angle-90:2*\Length);
  \draw[front1] (nu2) ++(\Angle+90:\Length) -- ++(\Angle-90:2*\Length);

  \draw[arc] (nu1) ++(0:\ArcRad) arc [start angle=0, end angle=\Angle, radius=\ArcRad] node[textstyle,midway,anchor=west,yshift=1mm] {$\theta$};

\end{tikzpicture}}
	\resizebox{0.57\linewidth}{!}{\tikzset{
  particlepath0/.style = {dotted,decoration={markings,mark=at position 0.50 with {\arrow[xshift=0.8mm]{stealth}}},postaction={decorate}},
  particlepath1/.style = {decoration={markings,mark=at position 0.50 with {\arrow[xshift=0.8mm]{stealth}}},postaction={decorate}},
  particlepath2/.style = {thin,dashed,opacity=0.6,decoration={markings,mark=at position 0.50 with {\arrow[xshift=0.8mm]{stealth}}},postaction={decorate}},
  front0/.style = {},
  front1/.style = {thin,dashed,opacity=0.6},
  arc/.style = {thin,opacity=0.6},
  textstyle/.style = {scale=1.3}
}

\begin{tikzpicture}[very thick]
  \def\Length{3cm}
  \def\Angle{45}
  \def\MarkerSize{1pt}
  \def\ArcRad{5mm}
  \def\La{0.8cm}

  \coordinate (start) at (0,0);
  \coordinate (nu1)   at (0,-\Length);

  \draw[color=black,domain=0:6,samples=100] plot ({3+1.4/exp(((\x-3)^2)/2)},-\x);
  \draw[front0] (start) ++(0:\Length) -- ++(-90:2*\Length);

  \draw[particlepath0] (start) ++(-90:\Length)   -- ++(0:\Length) coordinate (nu1) node[textstyle,midway,above] {$\nu$};
  \draw[particlepath1] (nu1) -- ++(\Angle:\Length) node[textstyle,right] {$\nu$};

  \draw[particlepath2] (start) ++(-90:\La) -- ++(0:\Length) coordinate (int);
  \draw[particlepath2] (int) -- ++(\Angle:\Length);
  \draw[particlepath2] (start) ++(-90:2*\Length-\La) -- ++(0:\Length) coordinate (int);
  \draw[particlepath2] (int) -- ++(\Angle:\Length);

  \draw[fill] (nu1) circle (\MarkerSize) node[below left,scale=1.2] {$\bm{N}_j$};
  \draw[fill] (nu1) circle (\MarkerSize) node[above left,scale=1.2] {$\bm{N}_k$};
  \draw[front1] (nu1) -- ++(0:2.5*\ArcRad);

  \draw[arc] (nu1) ++(0:\ArcRad) arc [start angle=0, end angle=\Angle, radius=\ArcRad] node[textstyle,midway,anchor=west,yshift=1mm] {$\theta$};

\end{tikzpicture}}
	\caption{Left panel: Front of incoming neutrino plane-wave (solid vertical line) scatters on nucleons at fixed positions, $\bm{x}_j$ and $\bm{x}_k$, respectively. 
		Non-zero scattering angle $\theta$ develops the phase difference $\Delta\varphi=\bm{q}(\bm{x}_j-\bm{x}_k)$ of two fronts of scattered neutrino plane-waves (dashed lines) which leads to a loss of coherence. 
		Right panel: Neutrino scatters off a $k$-th or $j$-th nucleon described by a wave-function exemplified here as a Gaussian profile.
		The outgoing neutrino wave, as  for any nucleon target, is a superposition of waves $e^{i\bm{q}\bm{x}_k}$ weighted by $\Big|\psi_n({\bm{x}_{1}}\dots {\bm{x}_{A}})\Big|^2$.
	}
	\label{fig:coh_flat}
\end{figure}
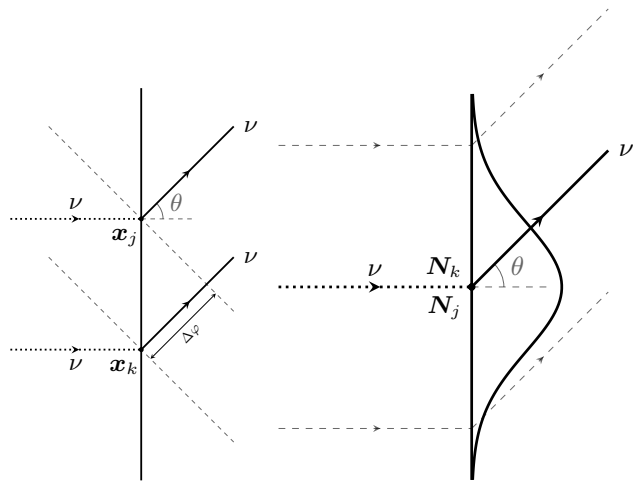

The left panel of~\cref{fig:coh_flat} depicts a neutrino scattering off of nucleons displaced from each other.
The non-zero  angle $\theta$ of the scattered neutrino leads to a loss of coherence. 

Does this consideration of coherency remain appropriate when the assumption of the nucleon's definite position is released?
In this case, the positions of nucleons are described by a multi-particle scalar wave-function $\psi_{n/m}({\bm{x}_{1}}\dots {\bm{x}_{A}})$, where the $n/m$ subscripts stand for the initial and final state of the nucleus.
The amplitude in~\cref{eq:amplitude_plane_wave} could be generalized as
\begin{equation}
\mathcal{A}_{nn} = \sum_{k=1}^A \mathcal{A}^k_{nn} f^k_{nn}(\bm{q}),
\label{eq:amplitude_wave_function}
\end{equation}
where 
\begin{equation}
\begin{aligned}
f^k_{mn}(\bm{q}) & =\langle m|e^{i\bm{q}\hat{\bm{X}}_k}|n\rangle\\
                            & = \int \Big(\prod_{i=1}^A d\bm{x}_i\Big) \psi_m^*({\bm{x}_{1}}\dots {\bm{x}_{A}})\psi_n({\bm{x}_{1}}\dots {\bm{x}_{A}})e^{i\bm{q}\bm{x}_k},\\
\end{aligned}
\label{eq:exp_generalize}
\end{equation}	
is the transition matrix element of  $e^{i\bm{q}\hat{\bm{X}}_k}$ with $\hat{\bm{X}}_k$ being the quantum position operator of the $k$-th nucleon.

In particular, 
\begin{equation}
f^k_{nn}(\bm{q})   = \int \Big(\prod_{i=1}^A d\bm{x}_i\Big) \Big|\psi_n({\bm{x}_{1}}\dots {\bm{x}_{A}})\Big|^2 e^{i\bm{q}\bm{x}_k},
\label{eq:form-factor}
\end{equation}
defining the form-factor of the nucleon  bound in the nucleus, differs from the exponential factor $e^{i\bm{q}\bm{x}_k}$ in two major respects.

(i) $f^k_{nn}(\bm{q})$ does not depend on the coordinate of the $k$-th nucleon. 
All position variables are integrated out in~\cref{eq:exp_generalize}.

(ii) $f^k_{nn}(\bm{q})$ does not depend on the index $k$ (ignoring for simplicity a possible difference in form-factors for protons and neutrons).
This statement can be proven for both fermions and bosons by a change of integration variables, and accounting for symmetry properties of the wave-function under interchange of its arguments.

Now, accounting for these properties of $f^k_{nn}(\bm{q})$ we conclude that phases of each individual amplitude in the total amplitude in~\cref{eq:amplitude_wave_function}  are all equal and the amplitudes are coherent for any $\bm{q}$, at variance with~\cref{eq:amplitude_plane_wave}.

This conclusion does not mean to say that the total amplitude is not vanishing at large $\bm{q}$, because in this limit the form-factor $f_{nn}(\bm{q})$ vanishes.
What governs such  dependence of $f_{nn}(\bm{q})$?
Mathematically, the reason lies in the fast oscillation of the $e^{i\bm{q}\bm{x}_k}$ factor in the integral in~\cref{eq:exp_generalize}, washing out the integrand function.
The physical reason is in the incoherent summation of waves belonging to the wave-function of a single nucleon extended over the size of the nucleus.
Other physics arguments are discusses in~\cref{sec:kinematic_paradox,app:mechanics}.

One can argue that this conclusion seems to be in conflict with a wave-function corresponding to the nucleons at fixed positions, assuming that 
\begin{equation}
|\psi_n({\bm{y}_{1}}\dots {\bm{y}_{A}})|^2 \propto \prod_{i}\delta^3(\bm{y}_i-\bm{x}_i), 
\label{eq:wf_fixed_positions_assumption}
\end{equation}
where $\bm{y}_i$ are variables and $\bm{x}_i$ are parameters.
Then,~\cref{eq:form-factor} reduces to~\cref{eq:amplitude_plane_wave} in which every term has an individual phase in contrast to our statement.
 This antinomy appeared because of the assumption in~\cref{eq:wf_fixed_positions_assumption} which breaks the principle of the particles identity.
The latter  requires that the multi-particle wave-function should be either symmetric (bosons) or anti-symmetric (fermions) under exchange of its arguments.
As a result it is not possible to state that the $i$-th particle has position $\bm{x}_i$  even if it is known that all particles occupy some fixed positions.
Instead, the $i$-th particle can be at any point among the $\bm{x}_1\dots {\bm{x}_A}$ fixed positions.
Therefore, considering~\cref{eq:form-factor} with an appropriately symmetrized $\delta$-like wave-function one would identically obtain~\cref{eq:amplitude_plane_wave} for any index $k$ in agreement with our conclusion.
Consideration of this antinomy is also helpful in understanding that the very form of~\cref{eq:amplitude_plane_wave} ignores the fundamental principle of quantum physics -- the indistinguishability of particles. 

The right panel of~\cref{fig:coh_flat} displays a scattering picture accounting for a wave-function of the nucleons exemplified here as a Gaussian profile.
The summation of waves weighted by $\Big|\psi_n({\bm{x}_{1}}\dots {\bm{x}_{A}})\Big|^2$ yields the scattered neutrino wave, as for any nucleon. 

Therefore, according to our consideration, it is not appropriate to identify the diagonal terms in 
\begin{equation}
\begin{aligned}
|\mathcal{A}_{nn}|^2 &= |f_{nn}(\bm{q})|^2\sum_{k,j}\mathcal{A}^k_{nn}\mathcal{A}^{j \, *}_{nn}\\
& = |f_{nn}(\bm{q})|^2\Big(\sum_k |\mathcal{A}^k_{nn}|^2 + \sum_{k\ne j} \mathcal{A}^k_{nn}\mathcal{A}^{j\, *}_{nn}\Big)
\end{aligned}
\end{equation}
as due to incoherent interactions.
Both diagonal and non-diagonal terms contribute equally to $|\mathcal{A}_{nn}|^2$, and with the same dependence on $\bm{q}$.

What, then, defines the incoherent interactions?
Essentially, they are defined by processes in which the quantum state of the nucleus is changed ($n\ne m$).
Let us briefly highlight the main points of a derivation illustrating this statement, ignoring for a while complications due to spin, type of nucleon, possible dependence of $\mathcal{A}_{mn}^k\to \mathcal{A}_{0}$ on the indices, etc (full details can be found in~\cref{app:cross_section}).

Assume the nucleus is initially in the $n$-th quantum state. 
If the experiment is not able to distinguish  the final state of the nucleus, one should sum over all possible final states to get the observable proportional to
\begin{equation}
\begin{aligned}
|\mathcal{A}|^2 &= \sum_{m}|\mathcal{A}_{mn}|^2 = |\mathcal{A}_{0}|^2\sum_{k,j}\sum_m f^k_{mn}f^{j\, *}_{mn}.
\end{aligned}
\label{eq:probability_1}
\end{equation}
Using~\cref{eq:exp_generalize} one can rewrite~\cref{eq:probability_1}  as
\begin{equation}
\begin{aligned}
|\mathcal{A}|^2 &= |\mathcal{A}_{0}|^2 \sum_{k,j}\langle n|e^{-i\bm{q}\hat{\bm{X}}_j}\sum_m|m\rangle\langle m|e^{i\bm{q}\hat{\bm{X}}_k}|n\rangle\\
&= |\mathcal{A}_{0}|^2 \sum_{k,j}\langle n|e^{-i\bm{q}\hat{\bm{X}}_j} e^{i\bm{q}\hat{\bm{X}}_k}|n\rangle,\\ 
\end{aligned}
\label{eq:probability_2}
\end{equation}
where we used the unity operator composed of nuclear states $\sum_m|m\rangle\langle m| = \hat{I}$.

One can define a two-particle real-valued correlation function
\begin{equation}
G_{nn}(\bm{q})  \equiv G(\bm{q}) = \langle n|e^{-i\bm{q}\hat{\bm{X}}_j} e^{i\bm{q}\hat{\bm{X}}_k}|n\rangle.
\label{eq:g_function}
\end{equation} 
If $k=j$, then $G(\bm{q})=1$.
For $k\ne j$, $G(\bm{q})$ does not depend on values of $k,j$ as can be seen using the symmetry properties of the nucleus wave-function.
Combining~\cref{eq:probability_2,eq:exp_generalize,eq:g_function} one gets
\begin{equation}
\begin{aligned}
|\mathcal{A}|^2 & = |\mathcal{A}_{0}|^2\left(A+G(\bm{q})A(A-1)\right)\\
& = |\mathcal{A}_{0}|^2\left(A^2G(\bm{q}) + A\left(1-G(\bm{q})\right)\right),\\
\end{aligned}
\label{eq:probability_3}
\end{equation}
where $A$ gives the number of nucleons.
The terms of $|\mathcal{A}|^2$ in~\cref{eq:probability_3}, quadratically  and linearly depending on $A$, are shaped by factors $G$ and $1-G$, respectively.
These terms provide a smooth transition between coherent and incoherent regimes.
One can observe that if the nucleus' multi-particle wave-function is constructed as a product of single-particle wave-functions, then $G(\bm{q})$ can be represented as $|F(\bm{q})|^2$, where $F(\bm{q})$ is the single-nucleon form-factor of the nucleus.

The derivation of~\cref{eq:probability_3} does not indicate in a transparent way what the source of the quadratic and linearly dependent terms is.
We conclude this section by showing that coherent and incoherent terms  are due to processes in which  the nucleus remains in the same quantum state or is changed, respectively.
For this purpose we rewrite~\cref{eq:probability_2} as
\begin{equation}
\begin{aligned}
|\mathcal{A}|^2 &= |\mathcal{A}_{0}|^2 \sum_{k,j}\langle n|e^{-i\bm{q}\hat{\bm{X}}_j}|n\rangle\langle n|e^{i\bm{q}\hat{\bm{X}}_k}|n\rangle\\
&+|\mathcal{A}_{0}|^2 \sum_{k,j}\sum_{m\ne n}\langle n|e^{-i\bm{q}\hat{\bm{X}}_j}|m\rangle\langle m|e^{i\bm{q}\hat{\bm{X}}_k}|n\rangle. 
\end{aligned}
\label{eq:probability_4}
\end{equation} 
The first line gives immediately
\begin{equation}
|\mathcal{A}_{0}|^2 A^2|F(\bm{q})|^2,
\label{eq:coherent_term}
\end{equation}
which could be identified as a coherent term in~\cref{eq:probability_3}.
The second line can be presented as
\begin{equation}
|\mathcal{A}_{0}|^2\left(A \left(1-|F(\bm{q})|^2\right) + \sum_{k\ne j}\text{cov}\left(e^{-i\bm{q}\hat{\bm{X}}_j},e^{i\bm{q}\hat{\bm{X}}_k}\right)\right),
\end{equation}
where the covariance of quantum operators reads
\begin{equation}
\begin{aligned}
\text{cov}\left(e^{-i\bm{q}\hat{\bm{X}}_j},e^{i\bm{q}\hat{\bm{X}}_k}\right)
&= \langle n|e^{-i\bm{q}\hat{\bm{X}}_j} e^{i\bm{q}\hat{\bm{X}}_k}|n\rangle \\
&- \langle n|e^{-i\bm{q}\hat{\bm{X}}_j}|n\rangle\langle n|e^{i\bm{q}\hat{\bm{X}}_k}|n\rangle.
\end{aligned}
\end{equation}
The covariance terms are identically zero for a multi-particle wave-function  constructed as a product of single-particle wave-functions and the second line of~\cref{eq:probability_4}
reads
\begin{equation}
|\mathcal{A}_{0}|^2A \left(1-|F(\bm{q})|^2\right).
\label{eq:incoherent_term}
\end{equation}
Therefore, one can conclude that an elastic process (first line in~\cref{eq:probability_4}) yields the coherent term in~\cref{eq:coherent_term}, while inelastic processes all together (second line in~\cref{eq:probability_4}) yield the incoherent term in~\cref{eq:incoherent_term}.

One can find a certain analogy with  the theory of neutrino oscillations in which the integration over an unobserved  time of neutrino emission leads to an incoherent  $L$-independent term in the oscillation probability formula (see, for example, in~\cite{Naumov:2017doctorthesis,Naumov:2010um}).

Attribution of elastic and inelastic processes as contributing to the coherent and incoherent interactions was also done in~\cite{Divari:2012zz,Pirinen:2018gsd,Tsakstara:2013lca,Donnelly:1975ze} where the authors performed numerical calculations of the corresponding cross-sections within appropriate nuclear models.
\subsection{Kinematics of elastic and inelastic neutrino-nucleus scattering}
\label{sec:kinematics}
In general, one should consider the treatment of neutrino-nucleus interactions using wave packets.
The corresponding formalism was developed (see, for example, Ref.~\cite{Karlovets:2016jrd,Naumov:2017doctorthesis}) and some potentially interesting effects for elastic neutrino-nucleus scattering could be envisaged and examined. 
We simplify our treatment by considering the initial and final states as having definite momenta.

Let us denote by $k=(E_\nu,\bm{k})$ and $k'=(E_\nu',\bm{k}')$ the four-momenta of incoming and outgoing neutrino, and by $P_n$ and $P'_m$ the four-momenta of initial and final state nuclei, respectively. 

The  total energy $P_n^0$ of a nucleus state $|P_n\rangle$ reads as $E_{\bm{P}}+\varepsilon_n$, where $\varepsilon_n$ is an internal energy of the nucleus state.
In the laboratory frame, energy $E_\nu'$ of the outgoing neutrino depends on  angle $\theta$ between $\bm{k}$ and $\bm{k}'$ 
\begin{equation}
\label{eq:outgoing_neutrino_energy_lab}
E_\nu' = \frac{m_A(E_\nu-\Delta\varepsilon_{mn})-E_\nu\Delta\varepsilon_{mn}+\Delta\varepsilon_{mn}^2/2}{m_A+E_\nu(1-\cos\theta)-\Delta\varepsilon_{mn}},
\end{equation}
where 
\begin{equation}
\label{eq:delta_Epsilon}
\Delta\varepsilon_{mn} = \varepsilon_m-\varepsilon_n
\end{equation}
is the difference of energies of the $|m\rangle$  and $|n\rangle$ states.
Absolute values of the four-momentum transfer vector, $q=(q_0,\bm{q})$, read
\begin{equation}
\begin{aligned}
q_0 & = E_\nu-E_\nu'=\Delta\varepsilon_{mn}+T_A,\\
|\bm{q}| & = \left(E_\nu^2+E_\nu^{'2}-2E_\nu E_\nu'\cos\theta\right)^{1/2}\simeq (2m_AT_A)^{1/2}, 
\end{aligned}
\end{equation}
where $T_A$ is the kinetic energy of the scattered nucleus, calculated below.

In neutrino-nucleus center-of-mass frame $q^2$ reads
\begin{equation}
q^2  = -4E_{\nu,n}^\star E_{\nu,m}^\star\sin^2\frac{\theta_\star}{2},
\label{eq:q2_cms}
\end{equation}
where 
\begin{equation}
E_{\nu,n}^\star  = \frac{s_{A,n}-m_{A,n}^2}{2\sqrt{s_A}}
\end{equation}
is the energy of neutrino scattering off a nucleon state $|n\rangle$,  $s_{A,n}=(k+P_n)^2$  and $m_{A,n}=m_A+\varepsilon_n$.

Minimum and maximum values of $q^2$ correspond to $\sin^2\frac{\theta_\star}{2}=1\text{ and } 0$, respectively
\begin{equation}
\label{eq:q2_bounds}
\begin{aligned}
q^2_\text{min} &= -4E_{\nu,n}^\star E_{\nu,m}^\star
                        = \frac{-4E_\nu^2}{\sqrt{\left(1+\frac{2E_\nu}{m_{A,n}}\right)\left(1+\frac{2E_\nu}{m_{A,m}}\right)}},\\
q^2_\text{max} &= 0.
\end{aligned}
\end{equation}

For heavy nuclei with $\Delta\varepsilon_{mn}$ of the order of hundreds keV and experimentally detectable signals produced by a release of kinetic energy of the scattered nucleus, $q^2$ can be approximated as
\begin{equation}
\label{eq:q2_estimate}
q^2 \approx -\bm{q}^2 \simeq -2m_AT_A.
\end{equation} 
Assuming the initial nucleus is at rest, the kinetic energy of its recoil reads
\begin{equation}
\label{eq:kinetic_energy_exact}
T_A = \sqrt{m_A^2+\bm{q}^2}-m_A.
\end{equation}
Using~\cref{eq:outgoing_neutrino_energy_lab} and assuming $m_A\gg E_\nu$, the kinetic energy $T_A$ of the scattered nucleus becomes
\begin{equation}
\label{eq:kinetic_energy}
T_A\approx \frac{E_\nu (E_\nu-\Delta\varepsilon_{mn})(1-\cos\theta)+\Delta\varepsilon_{mn}^2/2}{m_A}.
\end{equation} 
Here we will examine the kinetic energy for a few cases of interest.
(i) Forward scattering of the neutrino corresponds to $\cos\theta=1$ and yields the minimal kinetic energy of the nucleus
\begin{equation}
\label{eq:Tmin}
T_A^\text{min} = \lim_{\cos\theta\to1}T_A \approx \frac{\Delta\varepsilon_{mn}^2/2}{m_A},
\end{equation} 
which is zero for $m=n$ because no energy, nor three-momentum, is transferred in this case.
For $m\ne n$, the energy $q_0=\Delta\varepsilon_{mn}$ is transferred to the nucleus as well as the three-momentum $\bm{q}$, equal in magnitude to $q_0$ for forward scattering, thus yielding $T_A=\bm{q}^2/2m_A$.

(ii) Backward scattering  corresponds to $\cos\theta=-1$ and yields the maximal kinetic energy of the nucleus
\begin{equation}
\label{eq:TA_max}
T_A^\text{max} = \lim_{\cos\theta\to - 1}T_A \approx \frac{(2E_\nu -\Delta\varepsilon_{mn})^2}{2m_A},
\end{equation} 
which can be understood as follows.
For $m=n$ no energy is transferred to the nuclear structure, while the transferred three-momentum is equal to double the initial neutrino energy (backward scattering). %
Thus, $T_A=(2E_\nu)^2/2m_A$.
For  $m\ne n$, the energy $\Delta\varepsilon_{mn}$ transferred to the nucleus must be subtracted from the total transfered three-momentum $2E_\nu$, thus leading to~\cref{eq:TA_max}.

(iii) In general, the kinetic energy of the scattered nucleus is smaller if the nucleus changes its quantum state ($m\ne n$) with respect to the case when $m=n$.
Effectively, this can be described by a decrease of neutrino energy by $\Delta\varepsilon_{mn}$, which could be significant when $E_\nu$ and $\Delta\varepsilon_{mn}$ are comparable.

For heavy nuclei, like ${}^{133}\text{Cs}$ or ${}^{127}\text{I}$, used by the COHERENT experiment~\cite{Akimov:2017ade}, the first excitation energies are of the order of $\simeq 100$ keV, which are small corrections compared to the tens-of-MeV neutrino energies produced by the Spallation Neutron Source.
Therefore, the kinetic energy of the recoil nucleus is of the same order of magnitude for both elastic and inelastic scatterings.

\begin{figure}[!h]
	\includegraphics[width=\linewidth]{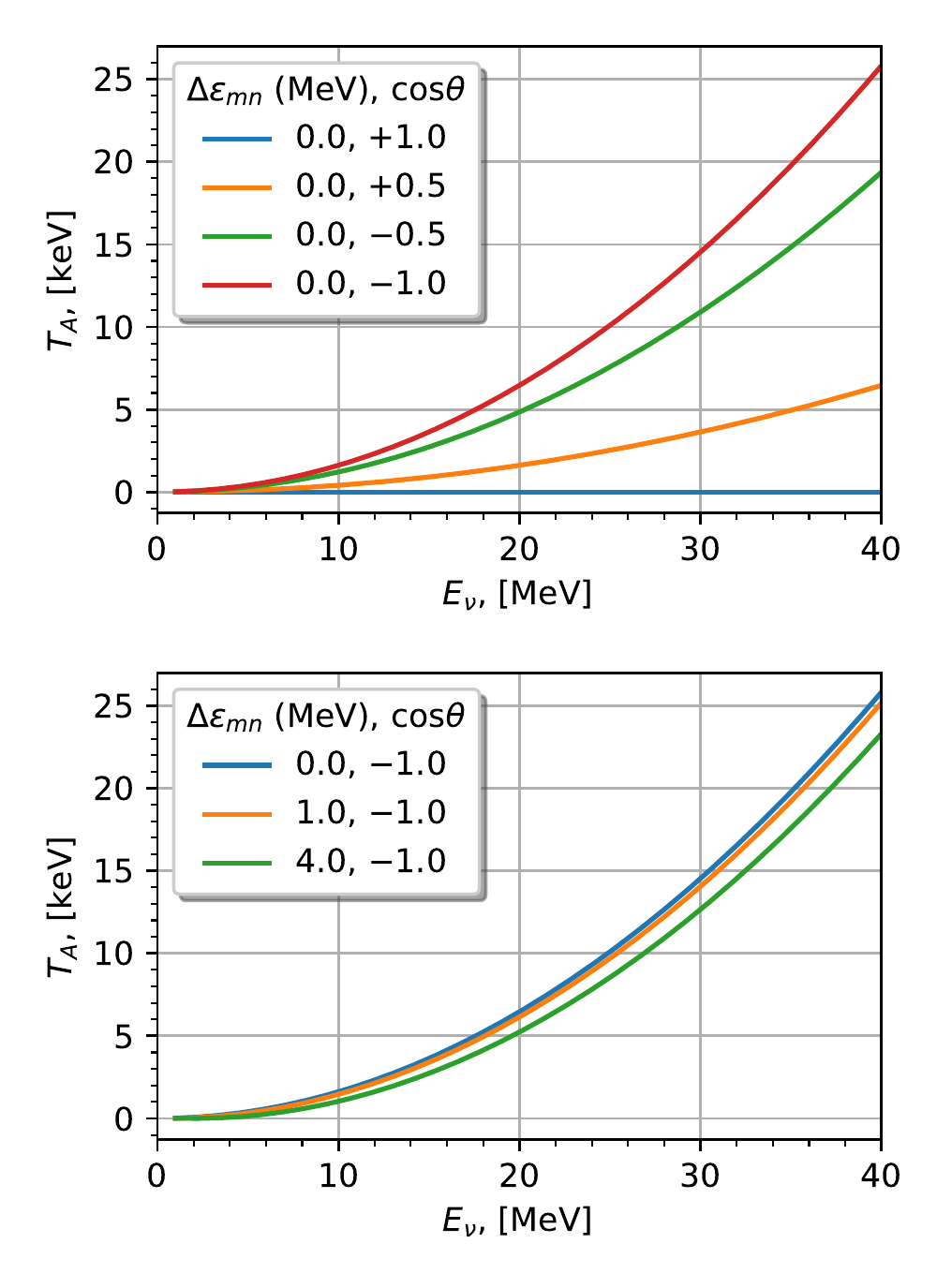}
	\caption{Expected kinetic energy of  nucleus ${}^{133}\text{Cs}$ scattered in an elastic interaction with a neutrino as a function of its energy. The upper plot corresponds to $\Delta\varepsilon_{mn}=0$ and four different values of $\cos\theta$, where $\theta$ is the neutrino scattering angle. The lower plot illustrates the impact of non-zero $\Delta\varepsilon_{mn}$ for a fixed value $\cos\theta=-1$.}
	\label{fig:kinetic_energy_nucleus}
\end{figure}

In~\cref{fig:kinetic_energy_nucleus} we show the expected kinetic energy of the recoil nucleus ${}^{133}\text{Cs}$  as a function of its kinetic energy, illustrating the impact of $\Delta\varepsilon_{mn}$ and $\cos\theta$. 
A strong dependence on neutrino scattering angle $\theta$ is evident from the upper panel of~\cref{fig:kinetic_energy_nucleus}.
The effect of non-zero values of $\Delta\varepsilon_{mn}$, displayed in the bottom panel of~\cref{fig:kinetic_energy_nucleus}, is also present, but it is significantly smaller than the angular dependence.
The reason for the weaker dependence due to non-zero values of $\Delta\varepsilon_{mn}$ is in the partial compensation due to the $\Delta\varepsilon^2_{mn}/2$ term in the numerator of~\cref{eq:kinetic_energy} at $E_\nu\simeq \Delta\varepsilon_{mn}$, and irrelevance of $\Delta\varepsilon_{mn}$ when $E_\nu\gg\Delta\varepsilon_{mn}$.

\subsection{Kinematic paradox}
\label{sec:kinematic_paradox}
As we show in what follows, the coherent enhancement of the interaction probability corresponds to neutrino-nucleus scattering in which the nucleus remains in the same quantum state.
This intuitively evident statement might seem to result in the following kinematic paradox.
Both the nucleon and the nucleus acquire the same three-momentum $\bm{q}$.
Assuming both of them are initially at rest, one arrives at the kinetic energy $T_N=\bm{q}^2/2m_N$ of the nucleon right after an interaction, which is a factor $m_A/m_N$ larger than the kinetic energy $T_A=\bm{q}^2/2m_A$ of the nucleus.
Since the nucleus remains in the same quantum state with the same internal energy,  $T_N$ and $T_A$ must be equal to each other, and to the difference $E_\nu-E_\nu^\prime$.

This paradox appears because some of the assumptions are incorrect.
In particular, while the assumption that the nucleon is at rest seems to be quite reasonable given that the average nucleon momentum, $\bm{p}$, is much smaller than  its mass, $m_N$, this  assumption  leads to the paradox.

\begin{figure}[!h]
	\resizebox{\linewidth}{!}{\tikzset{
  particlepath/.style = {decoration={markings,mark=at position 0.50 with {\arrow[xshift=1mm]{stealth}}},postaction={decorate}},
  momentum/.style     = {thick,decoration={markings,mark=at position 0.50 with {\arrow[xshift=1mm]{stealth}}},postaction={decorate}},
  boson/.style={decorate,decoration={snake}},
  nuclint/.style={thin,decorate,decoration={snake,amplitude=0.5mm,segment length=2mm}},
  secondary/.style={thin,opacity=0.55,text opacity=0.55},
  letters/.style={scale=1.5},
  smallletters/.style={scale=0.8}
}

\begin{tikzpicture}[very thick]
    \coordinate (start) at (0,0);
    \def\VecLength{20mm}
    \def\NucleusRadius{1.2*\VecLength}
    \def\NucleonRadius{3mm}
    \def\MomentumLength{10mm}
    \def\BosonDir{+0}
    \def\MomentumSpread{14}

    \def\CircleFrom#1#2#3#4#5{%
      \draw[#1] #2 ++(#3:#4) circle [radius=#4] coordinate #5;
    }
    \def\SnakeFromTo#1#2#3#4#5{%
      \draw[#1,nuclint] #2 -- ++(#3:#4) coordinate #5;
    }
    \def\SnakeFromToCircle#1#2#3#4#5#6{%
      \SnakeFromTo{#1,nuclint}{#2 ++(#3:#4)}{#3}{#5}{(save)}
      \CircleFrom{#1}{(save)}{#3}{#4}{#6}
    }
    \draw [particlepath] (start) -- ++(+60:\VecLength) coordinate (bos1) node [letters,midway,above,anchor=south east] {$\bm{k}$};
    \draw [particlepath] (bos1)  -- +(+120:\VecLength) node [letters,midway,above,anchor=north east] {$\bm{k'}$};
    \draw [boson,decoration={snake,segment length=3.0*\VecLength,amplitude=16mm}] (bos1)
      -- ++(\BosonDir:2*\VecLength)
      coordinate (bos2) node [letters,midway,below,anchor=north] {$\bm{q}=\bm{k}-\bm{k'}$} node[letters,midway,above,yshift=1.1cm] {$Z$};
    \draw (bos2) ++(+10:0.9*\VecLength) circle [radius=\NucleusRadius] coordinate (nucleus);

    \CircleFrom{}{(bos2)}{\BosonDir}{\NucleonRadius}{(nucleon)}
    \draw[momentum,dashed] (nucleon) ++(-\MomentumSpread+\BosonDir:\MomentumLength) coordinate (q1)
      -- ++(180-\MomentumSpread+\BosonDir:\MomentumLength) coordinate (last);
    \draw (q1) node[smallletters,anchor=west] {$\bm{p}=\alpha\bm{q}$};
    \draw[momentum] (last) -- ++(\MomentumSpread+\BosonDir:\MomentumLength)
      node[smallletters,anchor=west] {$\bm{p}+\bm{q}$};;

    \SnakeFromToCircle{secondary}{(nucleon)}{+80}{\NucleonRadius}{3*\NucleonRadius}{(nuclb)}
    \SnakeFromToCircle{secondary}{(nuclb)}{-30}{\NucleonRadius}{2.5*\NucleonRadius}{(nuclf)}
    \SnakeFromToCircle{secondary}{(nuclb)}{+10}{\NucleonRadius}{2.5*\NucleonRadius}{(nuclg)}
    \SnakeFromToCircle{secondary}{(nuclg)}{+10}{\NucleonRadius}{\NucleonRadius}{(nuclc)}
    \SnakeFromToCircle{secondary}{(nuclg)}{-40}{\NucleonRadius}{4.0*\NucleonRadius}{(nuclh)}
    \draw[secondary,nuclint] ($ (nuclf)!\NucleonRadius!(nuclh) $) -- ($ (nuclh)!\NucleonRadius!(nuclf) $);
    \SnakeFromToCircle{secondary}{(nucleon)}{-40}{\NucleonRadius}{3.5*\NucleonRadius}{(nucld)}
    \draw[secondary,nuclint] ($ (nucld)!\NucleonRadius!(nuclh) $) -- ($ (nuclh)!\NucleonRadius!(nucld) $);
    \SnakeFromToCircle{secondary}{(nucld)}{-170}{\NucleonRadius}{1.3*\NucleonRadius}{(nucle)}
    \SnakeFromToCircle{secondary}{(nucld)}{-30}{\NucleonRadius}{1.6*\NucleonRadius}{(nucli)}
    \SnakeFromToCircle{secondary}{(nucli)}{+50}{\NucleonRadius}{3*\NucleonRadius}{(nucli)}
    \draw[secondary,nuclint] ($ (nucli)!\NucleonRadius!(nuclh) $) -- ($ (nuclh)!\NucleonRadius!(nucli) $);

    \draw (nuclg) node (sump) [smallletters,above,secondary,yshift=1.5*\NucleonRadius] {$\bm{p}$};
    \draw[-stealth,secondary,dashed] ([xshift=-0.5*\MomentumLength]sump.north) -- ([xshift=0.5*\MomentumLength]sump.north);

    \draw[momentum] (nucleus) ++(\BosonDir:\NucleusRadius) -- ++(\BosonDir:\MomentumLength) node [letters,midway,below] {$\bm{q}$};
  \end{tikzpicture}}
	\caption{A qualitative picture of a coherent neutrino-nucleus interaction.
		A neutrino interacts with a nucleon initially having  a particular momentum $\bm{p}=\alpha\bm{q}$ aligned along $\bm{q}$ and given by~\cref{eq:nucleon_momentum_coherent}.
		Since the nucleus initially is at rest, all the nucleons except the target one have a momentum $-\bm{p}$ shown by line dashed.
		The final momentum  $\bm{p}+\bm{q} = (1+\alpha)\bm{q}$ of the target nucleon is also aligned along $\bm{q}$.
		In the figure an angle between  the $\bm{p}$ and $\bm{p}+\bm{q}$ vectors differs from $\pi$ for visual clarity.
		After the interaction  the increased energy of the target nucleon and acquired three-momentum $\bm{q}$ are transferred to the entire nucleus, leaving the  internal quantum state of the latter unchanged.
		A $Z$-boson having a wavelength comparable to the size of the nucleus produces a coherent enhancement of scattering amplitudes.}
	\label{fig:Multi-scattering-nuA}
\end{figure}
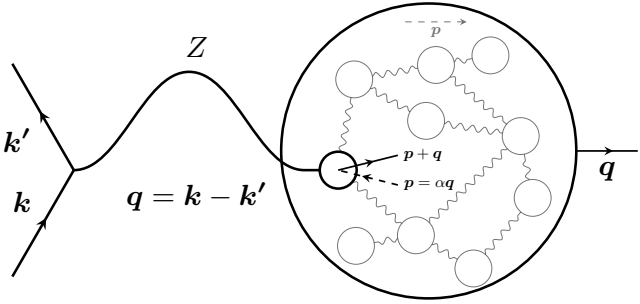

Let us require that $T_N$ and $T_A$ are equal to each other.
This requirement cannot be satisfied for any  nucleon momentum $\bm{p}$.
One can find a compatible $\bm{p}$ using energy conservation
\begin{equation}
\label{eq:energy_conservation_nucleon_nucleus1}
\frac{(\bm{p}+\bm{q})^2}{2m_N} -\frac{\bm{p}^2}{2m_N}= \frac{\bm{q}^2}{2m_A}.
\end{equation}
Searching for a solution where $\bm{p}$ is proportional to $\bm{q}$, $\bm{p}=\alpha \bm{q}$, one finds the nucleon momentum to be
\begin{equation}
\label{eq:nucleon_momentum_coherent}
\bm{p}= -\frac{\bm{q}}{2}\left(1-\frac{m_N}{m_A}\right).
\end{equation}
Therefore, energy-momentum conservation and the requirement that the nucleus does not change its state after an interaction provides a qualitative picture of the coherent neutrino-nucleus scattering process, displayed symbolically in~\cref{fig:Multi-scattering-nuA}. 
Here we discuss a few features of this interesting observation.
(i) Not every nucleon in the nucleus can interact with a neutrino in such a way that after the interaction the nucleus remains in the same state.
Only those nucleons which happen to have a momentum compatible with~\cref{eq:nucleon_momentum_coherent} are appropriate targets.

(ii) The wave-function of the nucleons provides us a distribution of the nucleon's momenta.
Large nucleon momenta are, in general, less probable than smaller momenta.
This explains qualitatively why at large $\bm{q}$ the enhancement factor in~\cref{eq:enhancement_factor_naive} vanishes, contrary to the case of small $\bm{q}$ for which the chance to find a nucleon with an appropriate momentum is relatively large.
Mathematically, this suppression is given by $|F(\bm{q})|^2$.

This consideration could be extended to the case of incoherent neutrino-nucleus scattering, when the nucleus changes its intrinsic quantum state $|n\rangle\to |m\rangle$ and $n\ne m$.
\cref{eq:energy_conservation_nucleon_nucleus1} must be generalized to account for non-zero differences of energy levels $\Delta\varepsilon_{mn}$
\begin{equation}
\label{eq:energy_conservation_nucleon_nucleus2}
\sqrt{E_{\bm{p}}^2+2\bm{p}\bm{q}+\bm{q}^2}-E_{\bm{p}}-\frac{\bm{q}^2}{2m_A}=\Delta\varepsilon,
\end{equation}
where $E_{\bm{p}}=\sqrt{m^2_N+\bm{p}^2}$. 
\cref{eq:energy_conservation_nucleon_nucleus2} does not use a non-relativistic approximation because for small values of $\bm{q}$, its solution $\bm{p}$ can be comparable to the nucleon mass.

\begin{figure}[!h]
	\includegraphics[width=\linewidth]{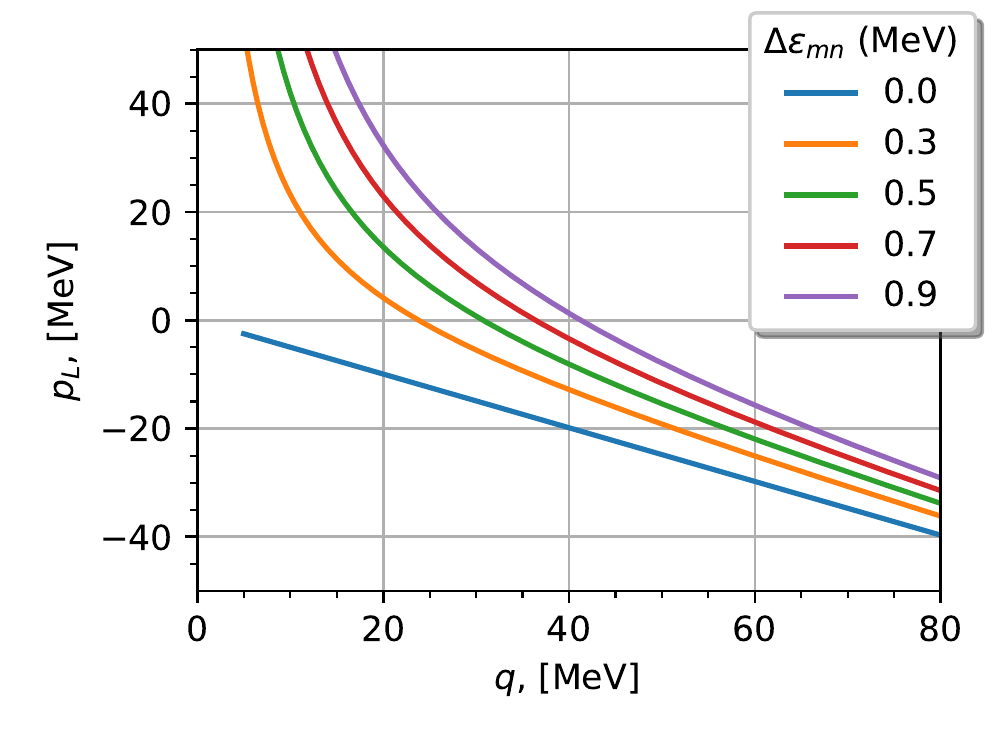}
	\caption{Longitudinal component $p_L$ of the nucleon momentum corresponding to the energy-momentum conservation in neutrino-nucleus scattering.}
	\label{fig:eff_nucleon_momentum}
\end{figure}

Splitting $\bm{p}$ into a sum of components: longitudinal $\bm{p}_L$ and transverse $\bm{p}_T$ to $\bm{q}$, one can find an exact solution of~\cref{eq:energy_conservation_nucleon_nucleus2}
\begin{equation}
 p_L = -\frac{|\mathbf{q}|}{2}\left(1-\sqrt{\beta}\sqrt{1+\frac{4m_{N,T}^2}{\mathbf{q}^2(1-\beta)}}\right),
 \label{eq:pL_solution_exact}
\end{equation}
where 
\begin{equation}
	\beta=\frac{E^2_{mn}}{\mathbf{q}^2}, \, E_{mn} = \frac{\mathbf{q}^2}{2m_A}+\Delta\varepsilon_{mn}, \, m_{N,T}^2 = m_N^2+p_T^2. 
	\label{eq:beta_etc}
\end{equation}
In~\cref{fig:eff_nucleon_momentum} we display the solution in~\cref{eq:pL_solution_exact} as a function of  $|\bm{q}|$.
For $\Delta\varepsilon_{mn}=0$ the solution in Fig.~\ref{fig:eff_nucleon_momentum} reproduces that in~\cref{eq:nucleon_momentum_coherent}.
In this case $\beta$ from~\cref{eq:beta_etc} is vanishingly small 
\begin{equation}
\beta = \frac{\bm{q}^2}{4m_A^2} = \frac{T_A}{2m_A}\ll 1
\end{equation}
and 
\begin{equation}
p_L \simeq -\frac{|\bm{q}_L|}{2}\left(1-\frac{m_{N,T}}{m_A}\right)
\end{equation}
coinciding with~\cref{eq:nucleon_momentum_coherent} for $p_T=0$.
One can observe that the longitudinal momentum of the nucleon in coherent neutrino-nucleus scattering is always aligned opposite to the transfered three-momentum $\bm{q}$.
For $\Delta\varepsilon_{mn}\ne 0$ the solution of~\cref{eq:energy_conservation_nucleon_nucleus2} is drastically different for small $\bm{q}$.
Here we analyze different regions of three-momentum transfer:
(i) At $\bm{q}=0$,~\cref{eq:energy_conservation_nucleon_nucleus2} has no solution, which simply means that at zero energy-momentum transfer an excitation of the nucleus is impossible.

(ii) At smallest $|\bm{q}|$ approaching its minimum possible value $|\bm{q}_\text{min}|=\Delta\varepsilon_{mn}+\displaystyle{\frac{\Delta\varepsilon_{mn}^2}{2m_A}}$,  the solution of~\cref{eq:pL_solution_exact} diverges, $p_L\to\infty$, whence a non-relativistic approximation in~\cref{eq:energy_conservation_nucleon_nucleus2} is not appropriate at small $\bm{q}$.
The chance to find a nucleon in the nucleus with such a momentum is vanishingly small.
Therefore, for small $|\bm{q}|$ the incoherent scattering is significantly suppressed, in opposition to  the coherent interaction. 

(iii) With increasing  $|\bm{q}|$ there is a good chance to find  a transition $|n\rangle\to |m\rangle$  with $\Delta\varepsilon_{mn}$  yielding $\bm{p}\approx 0$ for which the suppression is minimal. 
Again, this dependence is exactly opposite to the coherent scattering.

(iv) These kinematic considerations give a qualitative understanding, yet do not provide a complete picture of the dependence of the neutrino-nucleus scattering upon $\bm{q}$.
In $|n\rangle\to |m\rangle$ ($n\ne m$) transitions, the matrix element $\langle m|e^{i\bm{q}\hat{\bm{X}}}|n\rangle$, where $\hat{\bm{X}}$ is the position operator, determines the actual functional dependence.
A quantitative mathematical framework is developed in~\cref{app:amplitude}.

As a useful and simple illustration of transitions for which the internal state is changed or unchanged, in~\cref{app:mechanics} we  consider a mechanical analogy of a system of two balls with equal masses $m$ connected by a massless spring having non-zero rigidity.

\subsection{Scattering amplitude}
\label{sec:amplitude}
Our calculation follows a microscopic description of neutrino-nucleus scattering as a result of the neutrino-nucleon interaction.

We consider a Fock state $|P_n\rangle$ of a nucleus with four-momentum $P_n$ being in the $n$-th quantum state  as a superposition of free nucleons states weighted with their bound state wave-function.
The latter is explicitly factorized into a product of the wave-functions  describing the internal structure of the nucleus and motion of their center-of-mass.
The internal wave-function depends on $A-1$ three-momenta because one three-momentum variable is used to describe the motion of the nucleus.

It is convenient to refer to  the Fock state $|n\rangle$, describing the nucleus in the $n$-th quantum state at rest.
At zero nucleus momentum, both $|P_n\rangle$ and  $|n\rangle$ states describe the same quantum state but still differ by their normalizations given in~\cref{eq:nucleus_state_norm_1,eq:nucleus_state_norm_2}.
The details of this consideration are summarized in~\cref{app:framework}.

A priori, one does not know the initial, $|n\rangle$,  and final, $|m\rangle$, internal states of the nucleus.
Therefore,  all possible transitions must be considered.
The matrix element $i\mathcal{M}_{mn}$, corresponding to the process in~\cref{eq:neutrino_nucleus_scattering_definition} keeping only the leading order terms of Fermi constant $G_F$, reads
\begin{equation}
i\mathcal{M}_{mn}  = i\frac{G_F}{\sqrt{2}} \frac{m_A}{m_N}C_{1,mn}^{1/2} \sum_{k=1}^{A} \sum_{sr} f^k_{mn}\lambda^{mn}(s,r) (l,h^k_{sr}),
\label{eq:matrix_element}
\end{equation}
where $m_N$ and $m_A$ are masses of the nucleon and nucleus, respectively, and $C_{mn,1}$ is a function of the order of unity defined in~\cref{eq:C1_def}. 
Details of the derivation can be found in~\cref{app:amplitude}.

Functions  $f^k_{mn}(\bm{q})\equiv \langle m|e^{i\bm{q}\hat{\bm{X}}_{k}}|n\rangle$, where $\hat{\bm{X}}_{k}$ stands for the position operator of the $k$-th nucleon, are transition form-factors for $m\ne n$ and $n$-state form-factors for $m=n$, defined in~\cref{eq:hmunu_3}.

$(l,h^k_{sr})$ is the scalar product of the lepton $(l)$ and $k$-th nucleon's ($h^k_{sr}$) neutral weak currents.
For their definition refer to~\cref{eq:leptonic_current,eq:nucleon_current}, respectively.

$\lambda^{mn}(s,r)$ is a spin transition amplitude between the $|n\rangle$ and $|m\rangle$  states of the nucleus.
It depends on initial, $r$, and final, $s$, doubled spin projection on the given axis of the scattered nucleon. 
For a definition, refer to~\cref{eq:factorize_spin,eq:spin_functions_norm,eq:lambda_def}.

The  amplitude in~\cref{eq:matrix_element} is a sum  of  neutrino-nucleon amplitudes, each proportional to the scalar product of the lepton and nucleon currents, weighted by two factors, each not exceeding unity.

Given the definition of $f^k_{mn}$ in~\cref{eq:hmunu_3}  and the symmetry properties of the nucleus wave-function, one can conclude that $f^k_{mn}$ does not depend on the number $k$, but only on the type of nucleon $k$ points to.

Therefore, all amplitudes in~\cref{eq:matrix_element} have the same phase and thus are ''coherent'' in the literal sense of this terminology.

One can see that $f^k_{mn}(\bm{q})=\langle m|e^{i\bm{q}\hat{\bm{X}}_{k}}|n\rangle$ is a generalization of the $e^{i\bm{q} \bm{x}_k}$ quantum-mechanical factor used by Freedman in~\cite{Freedman:1973yd}.
\cref{sec:paradigm} can be referred for a discussion of an important difference between these two factors.

$f^k_{mn}$ is most important in understanding the mechanisms of the quadratic and linear dependence of the observable cross-section on the number of nucleons.

Let us examine the form-factor $f^k_{mn}$ for elastic and inelastic scatterings.

(i) In the case of elastic scattering
\begin{equation}
\lim_{\bm{q}\to 0}\langle n|e^{i\bm{q}\hat{\bm{X}}_{k}}|n\rangle\to 1
\end{equation}
and one expects  a quadratic dependence of  the cross-section on the number of nucleons. 
For $\bm{q}\to \infty$, the matrix element vanishes
\begin{equation}
\lim_{\bm{q}\to \infty}\langle n|e^{i\bm{q}\hat{\bm{X}}_{k}}|n\rangle\to 0
\end{equation}
and the elastic cross-section must also vanish.
Therefore, the elastic scattering has the properties of a ''coherent'' process in the terminology of Freedman.

(ii) In the case of inelastic scattering
\begin{equation}
\lim_{\bm{q}\to 0}\langle m|e^{i\bm{q}\hat{\bm{X}}_{k}}|n\rangle\to 0
\end{equation}
according to the normalization $\langle m|n\rangle=0$ for $n\ne m$ (see~\cref{eq:nucleus_state_norm_2}).
For a non-zero $\bm{q}$ the matrix element $\langle m|e^{i\bm{q}\hat{\bm{X}}_{k}}|n\rangle\ne 0$ in general, and as we show in~\cref{sec:cross_section} the cross-section  is a linear function of the number of nucleons once all possible initial and final states are accounted for.
Since this result could be obtained by summing up  the absolute values of the amplitudes squared one can refer to this case as incoherent scattering. 

\subsection{Cross-section}
\label{sec:cross_section}
The corresponding differential cross-section reads
\begin{equation}
\label{eq:cross-section_1}
\frac{d\sigma_{mn}}{dT_A} = \frac{|i\mathcal{M}_{mn}|^2}{2^5\pi E_\nu^2 m_A}C_{2,mn},
\end{equation}
where $C_{2,mn}$ is a function of the order of unity given in~\cref{eq:C2_mn}.
As we show in App.~\ref{app:matrix_element_calculation} (see~\cref{eq:scalar_products_sigma3})  the matrix-element squared, $|i\mathcal{M}_{mn}|^2$, is independent of the azimuthal angle $\varphi$, therefore we integrated over this variable in~\cref{eq:cross-section_1}.

An observable cross-section can be obtained by averaging over all possible initial states $|n\rangle$ and summing up over all possible final states $|m\rangle$
\begin{equation}
\label{eq:cross-section_2}
\frac{d\sigma}{dT_A} = \sum_{n,m}\omega_n \frac{d\sigma_{mn}}{dT_A},
\end{equation}
where $\omega_n$ is a statistical weight  to find an initial nucleus in a quantum state $|n\rangle$ at given ambient temperature. 
In what follows, we do not need an explicit form of $\omega_n$ normalized as
\begin{equation}
\label{eq:pn_normalization}
\sum_n \omega_n=1.
\end{equation}
The matrix-element squared, $|i\mathcal{M}_{mn}|^2$, has inside it a summation $\sum_{k,j}$ over two indexes  enumerating the scattered nucleons. 

In~\cref{app:cross-section} it is shown that terms in~\cref{eq:cross-section_2}, corresponding to elastic neutrino-nucleus scattering  ($\sum_{n=m}$), keep both indexes, $k$ and $j$, giving rise to a quadratic dependence of the cross-section as a function of the number of nucleons.
In contrast,  terms in~\cref{eq:cross-section_2}, corresponding to inelastic neutrino-nucleus scattering  ($\sum_{n\ne m}$), are to a good accuracy proportional to $\delta_{kj}$, which automatically yields a linear dependence on the cross-section as a function of the number of nucleons.

Therefore, the observable cross-section can be written as
\begin{equation}
\begin{aligned}
\frac{d\sigma}{dT_A} &= \frac{4G_F^2 m_A}{\pi}  \\
\times\Bigg(& g_i\sum_{f=n,p}\sum_{k=1}^{A_f} \sum_{s,r}|\lambda^f_{sr}|^2|(l,h^f_{sr})|^2(1-|F_f|^2)\\
+&g_c\left| \sum_{f=n,p}\sum_{k=1}^{A_f} \sum_r(l, h^f_{rr})F_f\right|^2\Bigg),
\end{aligned}
\label{eq:cross-section_3}
\end{equation}
where $|\lambda^{p/n}_{sr}|^2$  and $g_{i/c}$ are determined factorizing, respectively $|\lambda^{mn}_{sr}|^2$ given by~\cref{eq:spin_functions_norm2} and $g^{mn}$ defined by~\cref{eq:g_functions} out of the double sum $\sum_{nm}$ in~\cref{eq:cross-section_2}. 
$g_{i/c}$ are kinematic functions of the order of unity.
$|F_{p/n}|^2$ are proton and neutron form-factors of the nucleus  defined by~\cref{eq:hmunu_3}.

The second  and  third lines of~\cref{eq:cross-section_3} correspond to inelastic and elastic neutrino-nucleus scattering, respectively.
Their dependencies on the number of nucleons are linear and quadratic, respectively.
Using the terminology of Freedman, one would refer to these terms as incoherent and coherent, correspondingly.

This is the most general result of this work if  terms with covariances defined in~\cref{eq:cross_section_incoherent_term2,eq:cross_section_incoherent_term4} are neglected.

The summation of amplitudes due to the scattering off of various targets is evident in the third line of~\cref{eq:cross-section_3}.
Each type of nucleon is weighted according to the appropriate averaged form-factor $F_{p/n}(\bm{q})$.
Note, that the nucleus does not change its spin eigenstate in the coherent term.
This is encoded in the summation $\sum_r (l, h^{p/n}_{rr})$.

The incoherent term depends on  $|\lambda^{p/n}_{sr}|^2$.
The latter is a probability for a nucleon to change  spin index $r$ to $s$ in transitions $|n\rangle\to|m\rangle$, averaged  over  $n$ and summed up over $m$.

While one needs a model for the nucleus wave-functions to calculate  $|\lambda^{p/n}_{sr}|^2$, we approximate these coefficients by unity  $|\lambda^{p/n}_{sr}|^2\to 1$, which implies that for any $r$, any value of $s$ is possible with the same probability.
Therefore, we can complete our calculations of the cross-section.

The scalar products $(l,h^{p/n})$ are calculated in~\cref{app:matrix_element_calculation} using helicity and $\sigma_3$ bases.
The latter corresponds to the basis with spin projection on a fixed axis chosen to be along the incoming neutrino momentum.
While the results do not depend on the basis chosen, as demonstrated in~\cref{eq:relation_bases_squares}, it is more straightforward to use  the helicity basis with~\cref{eq:scalar_products_helicity} and $\sigma_3$ basis with~\cref{eq:scalar_products_sigma3} to calculate the incoherent and coherent cross-sections, respectively.

As follows from~\cref{eq:cross-section_3}, the observable neutrino-nucleus cross-section can be presented as a sum of incoherent and coherent   cross-sections
\begin{equation}
\label{eq:cross_section_4}
\frac{d\sigma}{dT_A} = \frac{d\sigma_\text{incoh}}{dT_A} + \frac{d\sigma_\text{coh}}{dT_A}.
\end{equation}
The incoherent cross-section reads
\begin{equation}
\label{eq:sigma-inc1}
\begin{aligned}
&\frac{d\sigma_\text{incoh}}{dT_A}   = \frac{2G_F^2 m_A}{\pi} g_i\sum_{f=n,p}\left(1-|F_f|^2\right)\\
&\times \Bigg[A_f \Bigg(g_{L,f}^2+g_{R,f}^2(1-y)^2-2g_{L,f}g_{R,f}\frac{y m_N^2}{s-m_N^2}\Bigg)\\
&+\,\,\Delta A_f \Bigg(\Big[g_{L,f}-g_{R,f}(1-y)\Big]\\
&\hspace{0.15\linewidth}\cdot\Big[g_{L,f}+g_{R,f}\Big(1-y\frac{s+m_N^2}{s-m_N^2}\Big)\Big]\Bigg)
\Bigg],
\end{aligned}
\end{equation}
where $g_L^{p/n}$ and  $g_R^{p/n}$ are left- and right-chirality couplings of the nucleons expressed via corresponding vector and axial couplings in~\cref{eq:left_right_couplings}.
The Bjorken $y$ is defined in~\cref{eq:Bjorken-y}.
The total energy squared $s=(p+k)^2$ of the neutrino and target nucleon is calculated assuming an effective momentum of the nucleon given by~\cref{eq:pL_solution_exact}.
Let us note that $y$ and $s$ are determined within neutrino-nucleon kinematics.

In~\cref{eq:sigma-inc1} $A_p=Z$, $A_n=N$ and $\Delta A_p\equiv \Delta Z=Z_+-Z_-$, $\Delta A_n\equiv \Delta N=N_+-N_-$, where $Z_\pm,N_\pm$ stand for the numbers of protons and neutrons with spin projection on the incident neutrino momentum axis equal to $\pm 1/2$.
A correction function $g_i^{p/n}$   of the order of unity is discussed earlier and is defined in~\cref{eq:g_functions}.

If the target nuclei are unpolarized, then terms  proportional to $\Delta A^f$ in~\cref{eq:sigma-inc1} vanish after averaging.
Therefore, for an unpolarized target the incoherent cross-section reads
\begin{equation}
\label{eq:sigma-inc2}
\begin{aligned}
&\frac{d\sigma_\text{incoh}}{dT_A}   = \frac{2G_F^2 m_A}{\pi} g_i\sum_{f=n,p}\left(1-|F_f|^2\right)\\
&\times A^f \Bigg(g_{L,f}^2+g_{R,f}^2(1-y)^2-2g_{L,f}g_{R,f}\frac{y m_N^2}{s-m_N^2}\Bigg).
\end{aligned}
\end{equation}
The coherent cross-section reads
\begin{equation}
\label{eq:sigma-coh1}
\begin{aligned}
&\frac{d\sigma_\text{coh}}{dT_A}   = \frac{G_F^2 m_A}{\pi}g_c\left(1-\frac{T_A}{T_A^\text{max}}\right)|G_V+G_A|^2,
\end{aligned}
\end{equation}
where 
\begin{equation}
\begin{aligned}
G_V & = \sum_f g_V^f F_f\Bigg(A_f\Big[1-\frac{y\tau}{2}\Big]+\Delta A_f\frac{y}{2}\Bigg),\\
G_A & = \sum_f g_A^f F_f\Bigg(\Delta A_f\Big[1-\frac{y}{2}\Big]+A_f\frac{y\tau}{2}\Bigg),
\end{aligned}
\label{eq:GV_GA1}
\end{equation}
where 
\begin{equation}
\tau = \frac{\sqrt{s}-m_N}{\sqrt{s}+m_N}.
\end{equation}
It is straightforward to perform the spin averaging in~\cref{eq:sigma-coh1}, removing the terms linear in $\Delta A_f$.
The final formula of the spin-averaged cross-section reads
 \begin{equation}
 \label{eq:sigma-coh2}
 \begin{aligned}
             &\frac{d\sigma_\text{coh}}{dT_A}   = \frac{G_F^2 m_A}{\pi}g_c\left(1-\frac{T_A}{T_A^\text{max}}\right)\sum_{f,f'}F_fF^*_{f'}\\
 \Bigg[ &g_V^fg_V^{f'}  \Bigg(A_f A_{f'} \left(1-\frac{y\tau}{2}\right)^2 + \Delta A_f \Delta A_{f'}\left(\frac{y}{2}\right)^2\Bigg)\\
           +&g_A^fg_A^{f'}  \Bigg(\Delta A_f \Delta A_{f'} \left(1-\frac{y}{2}\right)^2 +  A_f  A_{f'}\left(\frac{y\tau}{2}\right)^2\Bigg)\\
           +&2g_V^fg_A^{f'} \Bigg(A_fA_{f'}\left(1-\frac{y\tau}{2}\right)\frac{y\tau}{2}+\Delta A_f \Delta A_{f'}\frac{y}{2}\left(1-\frac{y}{2}\right)\Bigg].
 \end{aligned}
 \end{equation}
Finally,~\cref{eq:sigma-coh2} could be simplified if the following approximations are adopted.
(i) Terms proportional to $y\approx 3\%E_\nu/(30\text{MeV})$ are omitted.
(ii) Terms proportional to $\Delta A_f \Delta A_{f'}$ are neglected. 
This can  be done either for a spin-less nucleus, or approximately for heavy nuclei with $\Delta A\ll A$.
(iii) Terms proportional to $g_V^p$ are abandoned because $g_V^p\ll 1$.
(iv) The kinematic correction function $g_c\to 1$.

Therefore,  
\begin{equation}
\label{eq:sigma-coh3}
\frac{d\sigma_\text{coh}}{dT_A}   \approx \frac{G_F^2 m_A}{\pi}\left(1-\frac{T_A}{T_A^\text{max}}\right)|F_n|^2 \left(g_V^n\right)^2N^2, 
\end{equation} 
which is a well known result~\cite{Freedman:1973yd,Drukier:1983gj,Barranco:2005yy,Patton:2012jr,Papoulias:2015vxa,Smith:1985mta,Jachowicz:2001jr,Divari:2010zz,McLaughlin:2015xfa,Vergados:2009ei,Papavassiliou:2005cs,Divari:2012zz}.
Corrections to this formula are discussed in~\cref{sec:spin_axial}.

\section{Discussion}
\label{sec:discussion}
In what follows we discuss in detail the  calculated cross-section.
It is convenient to refer to the  cross-section integrated over the kinetic energy of the recoil nucleus
\begin{equation}
\label{eq:integral_xs}
\sigma(E_\nu)  = \int_{T_A^\text{min}} ^{T_A^\text{max}}\frac{d\sigma}{dT_A}dT_A.
\end{equation}
This integral depends on the energy threshold $T_A^\text{min}$, unique for each detector.
As  an illustration we consider three experimental setups.

We refer to the state-of-the-art energy thresholds of considered experimental setups, briefly described in what follows.

(i) A germanium detector exposed to $\overline{\nu}_e$ flux from a nuclear reactor.
Among all natural isotopes we select only one stable nucleus, ${}^{74}\text{Ge}$, for our illustration. 
The expected energy threshold for electrons of germanium bolometers is 200 eV~\cite{nuGEN}, which, accounting for the quenching in germanium detectors~\cite{Barker:2012ek}, roughly corresponds to 1 keV of the ${}^{74}\text{Ge}$ recoil kinetic energy.
We refer to the $\nu$GEN experiment at the Kalinin Nuclear Power Plant~\cite{Belov:2015ufh} as an example.
For illustration we calculate the differential cross-sections for two $\overline{\nu}_e$ energies, $5$ MeV and $7$ MeV, and total cross-section for $E_\nu\in (1,20)$ MeV.
As an estimate for an excitation energy of the ${}^{74}\text{Ge}$ nucleus we take $\Delta\varepsilon=900$ keV.

(ii) A CsI scintillator exposed to the neutrinos from the Spallation Neutron Source~\cite{Akimov:2017ade}.
The differential and   total cross-sections are calculated for $E_\nu=30$ MeV and $50$ MeV and for $E_\nu\in (1,150)$ MeV, respectively.
We assumed $\Delta\varepsilon=100$ keV for the ${}^{133}\text{Cs}$ nucleus.
The energy threshold was set to 5 keV of the ${}^{133}\text{Cs}$ recoil kinetic energy.

(iii) A liquid argon detector with an unprecedented low-energy threshold of $0.6$ keV for the ${}^{40}\text{Ar}$ nucleus achieved by the DarkSide Collaboration~\cite{Agnes:2018ves}.
The differential and   total cross-sections are calculated for $E_\nu=15$ MeV and for $E_\nu\in (1,50)$ MeV, respectively.

To make a prediction for an experiment we use (i) two form-factors $F_{p/n}(\bm{q})$ for protons and neutrons, respectively, and (ii) data regarding the energy levels of the nucleus under consideration.

We considered two models of the form-factors: symmetrized Fermi-distribution~\cite{Piekarewicz:2016vbn} and Helm form-factor~\cite{PhysRev.104.1466}. 
Both models of the form-factors give very similar results numerically if the parameters of the models are selected to reproduce the same proton and neutron RMS radii.
In what follows we present the results obtained assuming the same RMS radii for protons and neutrons, and using the Helm form-factors for definiteness. 

\begin{figure}[!h]
	\includegraphics[width=\linewidth]{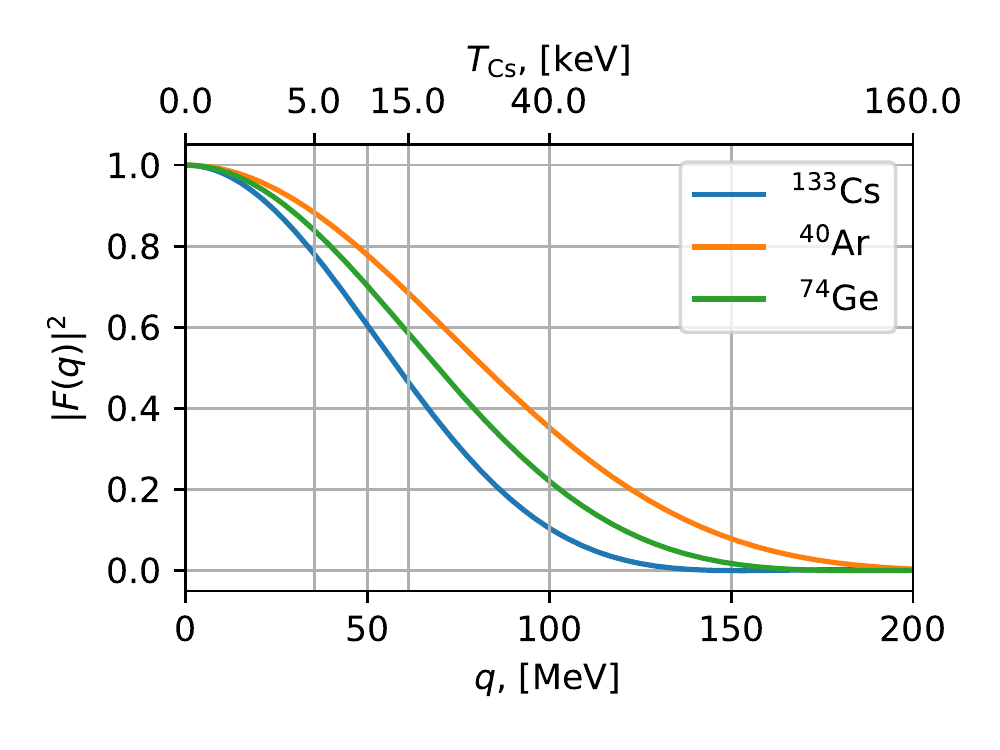}
	\caption{The Helm form-factor $F^\text{Helm}$~\cite{PhysRev.104.1466} as  a function of the absolute value of three-momentum transfer $|\bm{q}|$ (bottom horizontal axis).
	The upper horizontal axis corresponds to the kinetic energy of ${}^{133}\text{Cs}$ nucleus.}
	\label{fig:form-factors}
\end{figure}

In Fig.~\ref{fig:form-factors} predictions of these  models as functions of $|\bm{q}|$  are depicted.
At $T_A\simeq (12-15)$ keV, where the maximum of the signal observed by the COHERENT experiment occurred, $|\bm{q}|\simeq (50-60)$ MeV and $|F(\bm{q})|^2\simeq (0.6-0.5)$, indicating that pure coherent scattering has a suppression and a contribution from the incoherent transitions should be expected.

\subsection{Coherent and incoherent}
\label{sec:coherent_incoherent}
The most general feature of~\cref{eq:cross_section_4} consists of smooth transitions between coherent and incoherent regimes.
Both terms of the cross-section are governed by the same $F_{p/n}(\bm{q})$ form-factors defined in~\cref{eq:form_factors_1}.

In the limit $\bm{q}\to 0$, $F_{p/n}(\bm{q}) \to  1$, and the contribution of the incoherent cross-section  vanishes, while the coherent term totally dominates.

In the opposite limit of large $\bm{q}$, when $F_{p/n}(\bm{q}) \to 0$, the coherent cross-section vanishes and the incoherent term dominates.
In general, both coherent and incoherent scatterings contribute.

\begin{figure}[!h]
	\includegraphics[width=\linewidth]{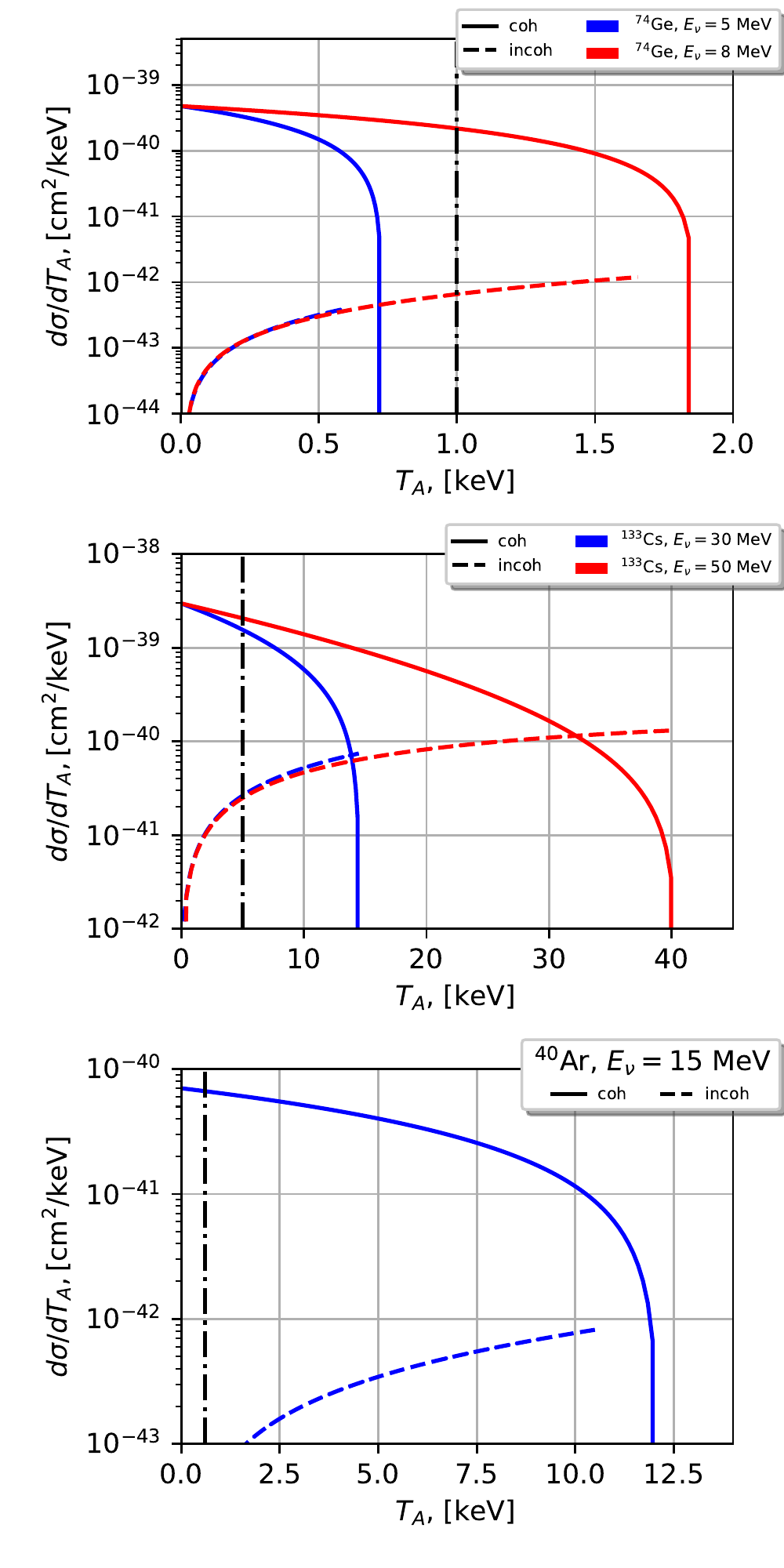}
	\caption{Differential cross-sections $\displaystyle{\frac{d\sigma}{dT_A}}$ for coherent (solid lines) and incoherent (dashed lines) neutrino-nucleus scattering for  ${}^{74}\text{Ge}$  (top), ${}^{133}\text{Cs}$ (middle) and ${}^{40}\text{Ar}$ (bottom) nuclei and different values of neutrino energies.
	Vertical lines correspond to experimental energy thresholds.
	}
	\label{fig:Xsection_differential}
\end{figure}

In~\cref{fig:Xsection_differential} the differential coherent and incoherent cross-sections are displayed for three experimental setups discussed above.

(i) At $T_A\to 0$ the coherent cross-section totally dominates since the incoherent contribution vanishes. 
For a given nucleus, the coherent differential cross-section in this limit does not depend on neutrino energy up to small corrections, in agreement with~\cref{eq:sigma-coh1}.

(ii) At $T_A\to T_A^\text{max}$ the coherent cross-section vanishes because of the factor $1-T_A/T_A^\text{max}$, while the incoherent cross-section rises. 
One might observe that the maximum kinetic energy of the nucleus experienced in an incoherent scattering is systematically smaller than that for the coherent interaction.
This is because some of the neutrino energy is used for the excitation of the nucleus, as given by~\cref{eq:TA_max}.

(iii) For small neutrino energies the coherent cross-section dominates over the incoherent contribution for any $T_A$.
For larger $E_\nu$ there is a value of $T_A$ above which the incoherent cross-section dominates over the coherent, as can be seen in the middle panel of~\cref{fig:Xsection_differential} for $E_\nu=50$ MeV.
In particular, for $E_\nu=50$ MeV with a ${}^{133}\text{Cs}$ nucleus this occurs at $T_A\gtrsim 33$ keV.

\begin{figure}[!h]
	\includegraphics[width=\linewidth]{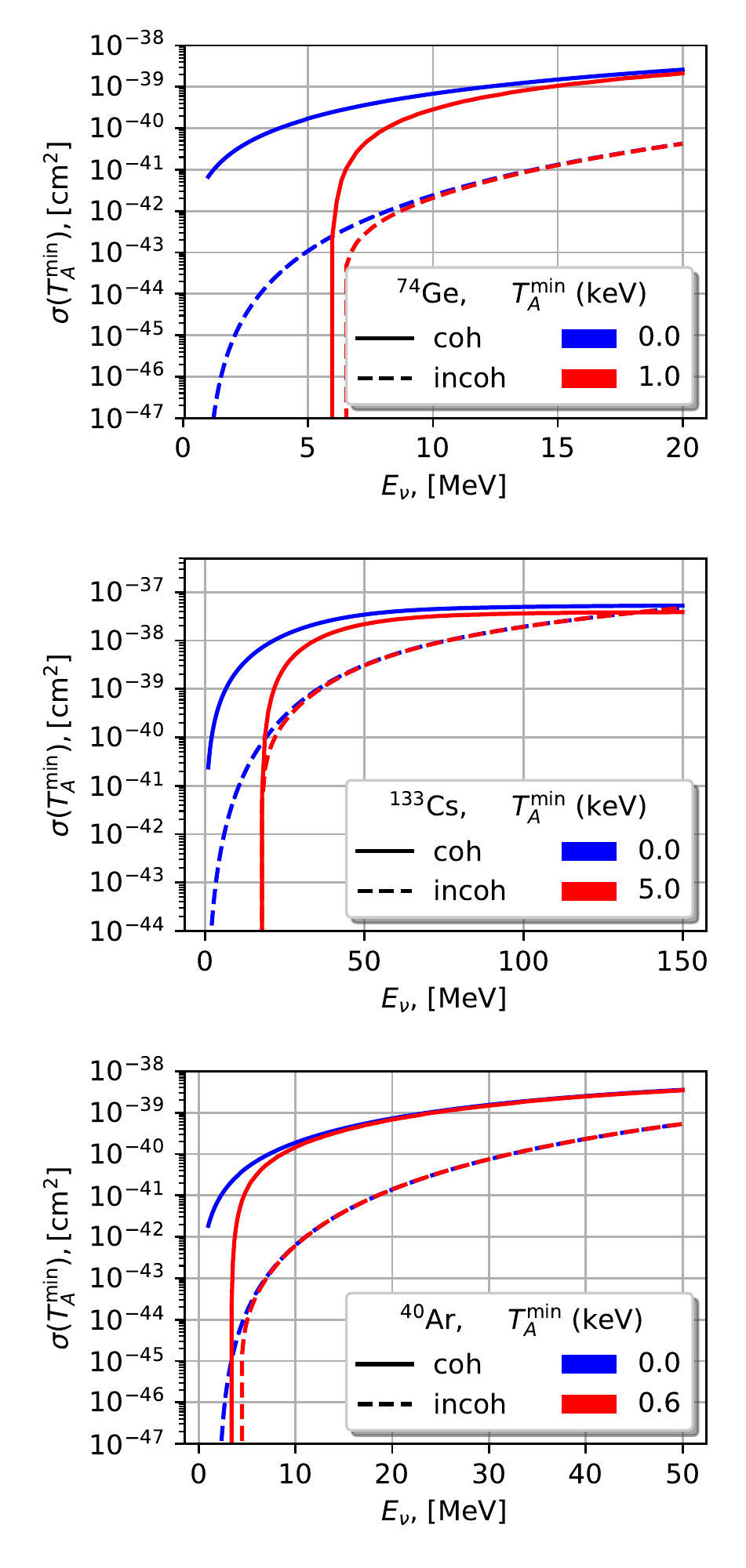}
	\caption{Integral cross-sections $\sigma$ for coherent (solid lines) and incoherent (dashed lines) neutrino-nucleus scattering for  ${}^{74}\text{Ge}$  (top), ${}^{133}\text{Cs}$ (middle) and ${}^{40}\text{Ar}$ (bottom) nuclei and different values of neutrino energies.
	The integrals are calculated for idealistic threshold-less ($T_A^\text{min}=0$, blue lines) experimental setups and accounting for state-of-the-art thresholds $T_A^\text{min}$ (red lines) achieved by three considered experimental setups.
	}
	\label{fig:Xsection_integral}
\end{figure}

In~\cref{fig:Xsection_integral} the corresponding integral cross-sections are displayed.

(i) At low $E_\nu$ the coherent integral cross-section is larger than the incoherent by orders of magnitude because the factors $1-|F_{p/n}(\bm{q})|^2$ suppress the latter at small $\bm{q}$.
With increasing neutrino energy their interrelation changes to the exact opposite, the incoherent cross-section dominating above a certain $E_\nu$.
As an example, for the ${}^{133}\text{Cs}$ nucleus this occurs at $E_\nu \gtrsim 140 \ (120)$ MeV for $T_A^\text{min}=0 \ (5)$ keV.

\begin{figure}[!h]
	\includegraphics[width=\linewidth]{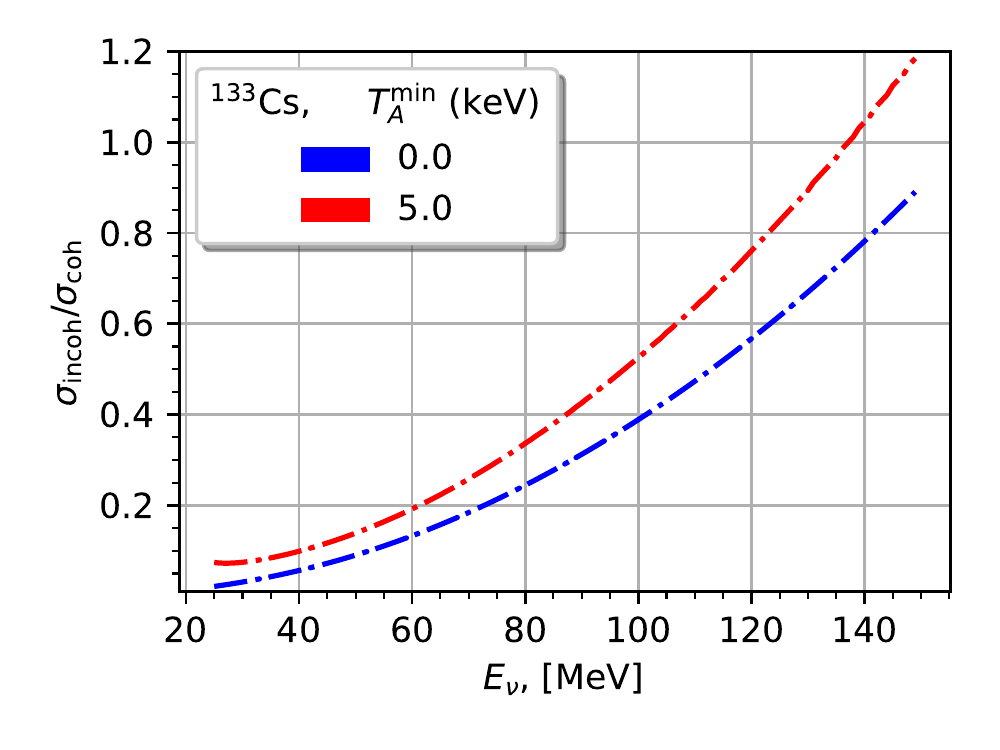}
	\caption{Ratio $\displaystyle{\sigma_\text{incoh}/\sigma_\text{coh}}$ for  neutrino scattering off of a ${}^{133}\text{Cs}$ nucleus as a function of $E_\nu$. The two curves correspond to a $T_A^\text{min}=0 \ (5)$ keV detection threshold.}
	\label{fig:Xsection_integral_ratios}
\end{figure}

(ii) The experimental detection threshold reduces the integrated coherent cross-section and, to a lesser extent the incoherent, because the threshold removes the part of the differential cross-section which is the largest for the former and vanishing for the latter, as can be seen in~\cref{fig:Xsection_differential}.
To quantify this statement the ratio of integrals given by~\cref{eq:integral_xs}, $\sigma_\text{incoh}/\sigma_\text{coh}$, is displayed in~\cref{fig:Xsection_integral_ratios} for the ${}^{133}\text{Cs}$ nucleus.
At $E_\nu\simeq 30 \ (50)$ MeV this ratio is about 5 (10)\% for an idealistic threshold-less experiment, and reaches about 10 (20)\% for $T_A^\text{min}=5$ keV.
The increasing importance of the incoherent interaction is evident for increasing neutrino energy.

\subsection{Revising the coherent cross-section}
\label{sec:spin_axial}
It is instructive to compare the coherent cross-section in~\cref{eq:sigma-coh1,eq:sigma-coh2} to that used in the literature~\cite{Akimov:2018ghi}
\begin{equation}
\begin{aligned}
\frac{d\sigma^0_\text{coh}}{dT_A} && = \frac{G_F^2 m_A}{2\pi}\Bigg[&\left(G_V+G_A\right)^2+\left(G_V-G_A\right)^2(1-y)^2\\
																&&													 &-\left(G_V^2-G_A^2\right)\frac{m_AT_A}{E_\nu^2}\Bigg]\\
																&&\approx \frac{G_F^2 m_A}{\pi}\Bigg[&G_V^2\left(1-\frac{T_A}{T_A^\text{max}}\right) + G_A^2\left(1+\frac{T_A}{T_A^\text{max}}\right)\Bigg]\\
																&&\approx \frac{G_F^2 m_A}{\pi}&\left(1-\frac{T_A}{T_A^\text{max}}\right)|F_n|^2(g_V^n)^2N^2,
\end{aligned}
\label{eq:coh_cross_section_ref}
\end{equation}
where $G_V = F_pZg_V^p + F_nNg_V^n$ and $G_A=F_p\Delta Z g_A^p + F_n\Delta N g_A^n$.

The second approximate equality of~\cref{eq:coh_cross_section_ref} appeared as a result of a quite accurate approximation $y = T_A/E_\nu \to 0$.
The last line is a result of further approximations: (i) $g_V^p\to 0$ and (ii) spin-less nucleus.

Let us briefly review~\cref{eq:coh_cross_section_ref}.
After a number of approximations, the third line of~\cref{eq:coh_cross_section_ref} is identical to an approximation of the coherent cross-section in~\cref{eq:sigma-coh3}, calculated in this work.
However, conceptually, a derivation of~\cref{eq:coh_cross_section_ref} is at odds with the coherency. 
Indeed, as one can observe, the first line of~\cref{eq:coh_cross_section_ref} corresponds to a calculation of incoherent cross-section (compare to~\cref{eq:sigma-inc2}), where the nucleus changes its spin eigenstate.
As we advocate here, the coherent scattering corresponds to interactions of neutrino with the nucleon in which the latter remains in the same quantum state.

How then are~\cref{eq:coh_cross_section_ref} and ~\cref{eq:sigma-inc1}  consistent with a good accuracy?
The reason is in the non-relativistic approximation.
Two terms of the matrix element containing $(l,h^\eta_{+-})$ and $(l,h^\eta_{-+})$ with a spin-flip should not contribute to the coherent cross-section (on the opposite, they do exist in~\cref{eq:coh_cross_section_ref}).
In the non-relativistic approximation $(l,h^\eta_{+-})$ vanishes, while $(l,h^\eta_{-+})$ is proportional to $g_A$ and vanishes for a spin-less nucleus, as can be seen in~\cref{eq:scalar_products_sigma4}.
The last statement is accurate if the nucleons in the nucleus are at rest.

\begin{figure}[!h]
	\includegraphics[width=\linewidth]{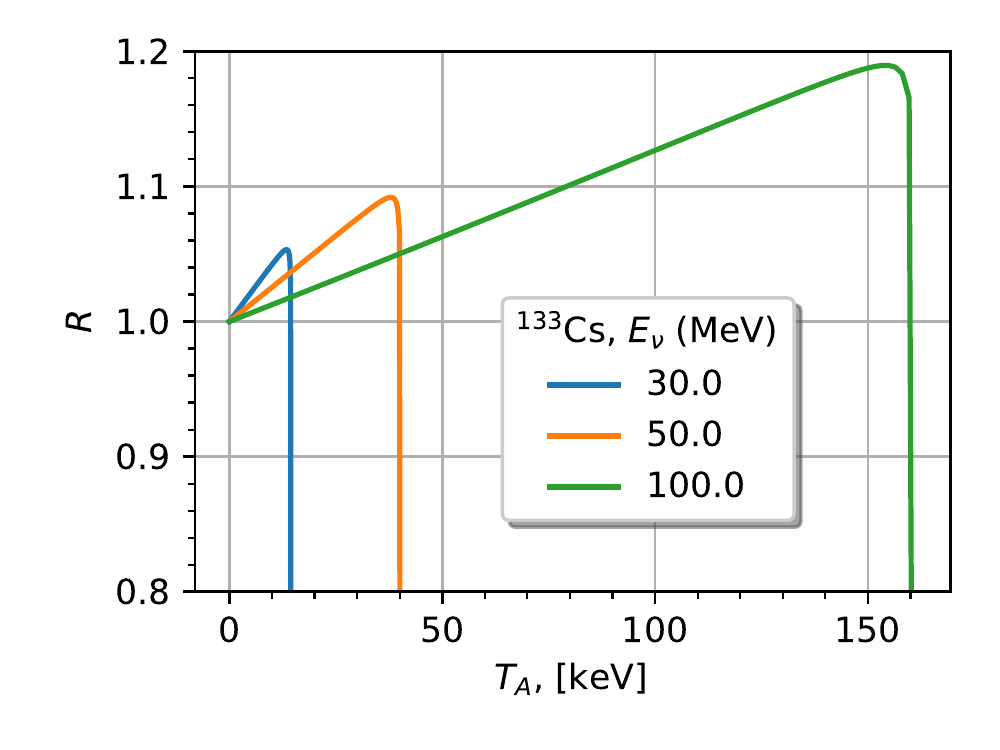}
	\caption{Ratio $R$ of differential coherent cross-section $d\sigma/dT_A$ calculated in this work in~\cref{eq:sigma-coh1} to that used by the COHERENT Collaboration and reproduced in~\cref{eq:coh_cross_section_ref}.
		Both cross-sections are averaged over the nucleus spin, assuming  neutrino scattering off of a ${}^{133}\text{Cs}$ nucleus. 
		The ratio is shown  as a function of kinetic energy  $T_A$ of the nucleus. }
	\label{fig:cross-section-ratio-ref}
\end{figure}

To illustrate the effects of a moving target nucleon and constant spin of the nucleus in elastic neutrino-nucleus scattering, a ratio of differential coherent cross-section $d\sigma/dT_A$  in~\cref{eq:sigma-coh1} to that  in~\cref{eq:coh_cross_section_ref} is displayed in~\cref{fig:cross-section-ratio-ref} for a ${}^{133}\text{Cs}$ nucleus, assuming three fixed values of neutrino energy.
The cross-sections coincide at $T_A=0$ and show a difference at some percent with increasing $T_A$.
The maximal difference occurring at the end of the nucleus kinetic energy spectra, rises with neutrino energy from about 5\% at $E_\nu=30$ MeV to about 20\% at $E_\nu=100$ MeV.

\subsection{Proposal to observe higher energy excitation $\gamma$s due to incoherent scattering}
\label{sec:gamma_proposal}
After an interaction the nucleus may remain in the same quantum state, or the internal  state of the nucleus could be changed. 
We refer to these cases as elastic and inelastic interactions.
Experimentally, the scattered nucleus, being in the same or  an excited state, are practically indistinguishable if one measures only the kinetic energy of the nucleus.
Inelastic interactions must be accompanied by the emission of gammas corresponding to the difference of energy levels of the nucleus.
The time scale  of these emissions is in the range of picoseconds to nanoseconds for the ${}^{133}\text{Cs}$ nucleus, taken as an example.
The energies of the $\gamma$s are of the order of some hundred keV for the same nucleus.
These $\gamma$s should produce a very detectable signal in the scintillator correlated in time with the beam pulses for an accelerator based experiment.
The rate of these $\gamma$s is determined by the ratio $N_\text{inc}/N_\text{coh}$, where
\begin{equation}
N_\text{incoh/coh} = \int dE_\nu \Phi(E_\nu)\int_{dT_A^\text{min}}^{dT_A^\text{max}} dT_A\frac{d\sigma_\text{inc/coh}}{dT_A} \varepsilon(T_A),
\end{equation}
in which $\varepsilon(T_A)$ is the detection efficiency.
\cref{fig:Xsection_integral_ratios} suggests that the number of $\gamma$ events due to incoherent interactions  should be detectable.

It is remarkable, that a similar proposal was made back to 1975 in~\cite{Donnelly:1975ze}.
\section{Summary}
\label{sec:summary}
A theoretical framework for neutrino-nucleus scattering $\nu A\to \nu A$, in which the nucleus conserves its integrity, is developed.
The main result of this work consists in the demonstration that coherent and incoherent regimes appear due to elastic and inelastic processes, when all possible initial and final states are taken into account.
This conclusion is in agreement with  corresponding theories of scattering of $X$-rays, electrons of an atom and of slow neutrons off  matter constituents.

The coherent and incoherent cross-sections were shown to be driven by $|F_{p/n}|^2$ and $(1-|F_{p/n}|^2)$ factors, thus providing a smooth transition between these regimes.

We also revised a formula for the coherent cross-section. 
The obtained formula  has some percent level corrections when compared to that known in the literature (see, for example in ~\cite{Akimov:2018ghi}).
They differ at most at the end of kinetic energy spectrum of the target nucleus, reaching   $\approx 5$\% at $E_\nu=30$ MeV ($\approx 20$\% at $E_\nu=100$ MeV).
There are two main sources for this difference.
(i) Our consideration treats only those matrix elements which correspond to the same initial and final  spin states of the nucleus in contrast to the conventional derivation which considers also the spin-flipped matrix elements.
(ii) The target nucleon is not assumed to be at rest which develops corrections to the vector and axial form-factors of the nucleus.

Three experimental setups considered in this work illustrate our results.
In particular, for ${}^{133}\text{Cs}$ and neutrino energies of $30-50$ MeV the incoherent cross-section is about 10-20\% of the coherent contribution if experimental detection threshold is accounted for.
The incoherent processes being a relatively small ''background'' to the coherent interactions provide an important clue if $\gamma$s released by excited nucleus are detected.
Detection of both signals due to nuclear recoil and excitations $\gamma$s provides a more sensitive instrument in studies of nuclear structure and possible signs of new physics.

An interested reader could checkout and run a Jupyter Notebook where equations from this manuscript are documented in terms of a python code~\cite{NNCoherence_Jupyter}.
\begin{acknowledgments}

We are grateful to D.~Dwyer, C.~Giunti, M.~Gonchar, S.~E.~Korenblit, C.~Kullenberg, V.~A.~Kuzmin, J.~Link, V.~A.~Naumov,  O.~Yu.~Smirnov, E.~Yakushev, S.~Zhou, X.~Qian for reading the manuscript and making important comments.

\end{acknowledgments}

\appendix
\section{Decomposition of $n$-particles states in $x$ and $p$ bases}
\label{app:framework}
In this section we shortly summarize some mathematical  aspects of the  representations of abstract quantum states for both single fermion and $n$-fermions.
\subsection{Single-particle states}
\label{app:one_particle_states}
We begin by reminding the reader about the single-particle basis.
%
A fermionic state with mass $m$, definite three-momentum $\bm{p}$, energy $E_{\bm{p}}=\sqrt{\bm{p}^2+m^2}$ and spin projection $s$ is defined according to
\begin{equation}
|{\bm{p}},s\rangle = \sqrt{2E_{\bm{p}}} a^\dagger_{{\bm{p}},s}|0\rangle,
\label{eq:p_definition}
\end{equation}
with Lorentz-invariant normalization
\begin{equation}
\langle \bm{k},s|\bm{p},r\rangle = (2\pi)^32E_{\bm{p}}\delta^3(\bm{p}-\bm{k})\delta_{rs}.
\label{eq:1particle_state_norm}
\end{equation}
A fermionic state with definite $\bm{x}$ can be defined as
\begin{equation}
\langle\bm{x}| = \langle 0|\hat{\psi}(\bm{x}),
\label{eq:x_definition}
\end{equation}
where $\hat{\psi}(\bm{x})$ is the free field operator in the Schroedinger representation.
The state in~\cref{eq:x_definition} is a Dirac spinor. 
These states are normalized as follows
\begin{equation}
\langle\bm{y}|\bm{x}\rangle = \delta^3(\bm{x}-\bm{y})\hat{I}_{4\times 4},
\end{equation}
where $\hat{I}_{4\times 4}$ is the $4\times 4$ unity matrix in the spinor space.

The single-particle unity operators read
\begin{equation}
\begin{aligned}
\hat{I}_{p,1} &= \int \frac{d\bm{p}}{(2\pi)^3}\frac{|\bm{p},s\rangle\langle \bm{p},s|}{2E_{\bm{p}}},\\
\hat{I}_{x,1} &=\int d\bm{x} |\bm{x}\rangle\langle\bm{x}|.
\end{aligned}
\label{eq:unity_operators_1}
\end{equation}
The scalar product of states given by~\cref{eq:p_definition,eq:x_definition} 
\begin{equation}
\langle\bm{x}|\bm{p},s\rangle = u(\bm{p},s)e^{i\bm{p}\bm{x}}
\end{equation}
allows for the representation of $|\bm{p},s\rangle$ and $|\bm{x}\rangle$ states via linear superpositions of each other
\begin{equation}
\begin{aligned}
\langle\bm{p},s| & =  \int d\bm{x} \, u^\dagger(\bm{p},s)e^{-i\bm{p}\bm{x}} \langle\bm{x}|,\\
\langle\bm{x}|    & = \int\frac{d\bm{p}}{(2\pi)^3} \frac{u(\bm{p},s) e^{i\bm{p}\bm{x}}}{{2E_{\bm{p}}}} \langle\bm{p},s|.
\end{aligned}
\label{eq:x_p_via_each_other}
\end{equation}
The second line of~\cref{eq:x_p_via_each_other} allows us to see that $\langle\bm{x}| $, given by~\cref{eq:x_definition}, differs from a non-relativistic spin independent state
\begin{equation}
\langle\mathrm{\bf{x}}|  = \int \frac{d\bm{p}}{(2\pi)^3} \frac{e^{i\bm{x}\bm{p}}}{\sqrt{2E_{\bm{p}}}}\langle\bm{p}|,
\label{eq:x_definition_nonrel}
\end{equation}
where $\langle\bm{p}|$ is defined similarly to~\cref{eq:p_definition} but for a spin-less particle.

A one-particle state $|\psi\rangle$ can be represented through $|\bm{p},s\rangle$ and $|\bm{x}\rangle$ states
\begin{equation}
\begin{aligned}
|\psi\rangle  &\equiv \hat{I}_{p,1}|\psi\rangle =  \int \frac{d\bm{p}}{(2\pi)^3} |\bm{p},s\rangle\frac{\widetilde{\psi}(\bm{p},s)}{\sqrt{2E_{\bm{p}}}},\\
|\psi\rangle  &\equiv \hat{I}_{x,1}|\psi\rangle =  \int d\bm{x}   |\bm{x}\rangle\psi(\bm{x}) ,
\end{aligned}
\label{eq:state_representations_1dim}
\end{equation}
where $\widetilde{\psi}(\bm{p},s) = \langle\bm{p},s|\psi\rangle/\sqrt{2E_{\bm{p}}}$ and $\psi(\bm{x})=\langle\bm{x}|\psi\rangle$.

Imposing $\langle\psi|\psi\rangle=1$ the wave-functions $\widetilde{\psi}(\bm{p},s)$ and $\psi(\bm{x})$ are normalized according to
\begin{equation}
\int \frac{d\bm{p}}{(2\pi)^3} |\widetilde{\psi}(\bm{p},s)|^2 = \int d\bm{x}|\psi(\bm{x})|^2=1.
\label{eq:wave_function_normalization_1dim}
\end{equation}
These wave-functions are related to each other through the Fourier transform 
\begin{equation}
\begin{aligned}
\psi(\bm{x})                        & = \int \frac{d\bm{p}}{(2\pi)^3} \frac{\widetilde{\psi}(\bm{p},s)}{\sqrt{2E_{\bm{p}}}}  u(\bm{p},s)e^{i\bm{p}\bm{x}},\\
\widetilde{\psi}(\bm{p},s) & =\frac{u^\dagger(\bm{p},s)}{\sqrt{2E_{\bm{p}}}} \int d\bm{x}  \psi(\bm{x})e^{-i\bm{p}\bm{x}}.
\end{aligned}
\end{equation}
Note that $\widetilde{\psi}(\bm{p},s)$ is  a scalar, while $\psi(\bm{x})$ is a Dirac spinor.

\subsection{$n-$particle states}
The unity operators defined in~\cref{eq:unity_operators_1}, generalized for $n$-particle states, reads
\begin{equation}
\begin{aligned}
&\hat{I}_{p,n} = \int \left(\prod_{i=1}^n \frac{d\bm{p}_i}{(2\pi)^32E_{\bm{p}_i}}\right) \frac{|\{p\}\rangle\langle \{p\}|}{n!}, \\
&\hat{I}_{x,n} = \int \left(\prod_{i=1}^n d\bm{x}_i\right) \frac{|\{x\}\rangle\langle\{x\} |}{n!}.
\end{aligned}
\label{eq:unity_operators_n}
\end{equation}
The symbols $\{p\}$ and $\{x\}$ are $n$-tuples, $\{p\}=(p_1\dots p_n)$ and $\{x\}=(x_1\dots x_n)$ are used for compaction here and in what follows. 
The  bra-vector $\langle\{x\}|$ is given as
\begin{equation}
\langle {\{x\}}| = \langle 0|\psi_{m_1}(\bm{x}_1)\dots \psi_{m_n}(\bm{x}_n),
\label{eq:x_definition_n_particles}
\end{equation}
with $x_i=(\bm{x}_i,m_i)$, where $m_i$ enumerates the spinor's rows of the fields $\psi(\bm{x}_i)$. 

Similarly to~\cref{eq:state_representations_1dim} the wave-functions in both momentum and coordinate spaces for the $n$-particle state $|\psi\rangle$ can be obtained using $|\psi\rangle = \hat{I}_{p,n}|\psi\rangle=\hat{I}_{x,n}|\psi\rangle$
\begin{equation}
\begin{aligned}
|\psi\rangle & = \int \left(\prod_{i=1}^n d\widetilde{\bm{p}}_i\right)\frac{\widetilde{\psi}(\{p\})}{\sqrt{n!}} |\{p\}\rangle,\\
|\psi\rangle & = \int \left(\prod_{i=1}^n  d\bm{x}_i \right) \frac{\psi(\{x\})}{\sqrt{n!}}|\{x\}\rangle,
\end{aligned}
\end{equation}
where 
\begin{equation}
\begin{aligned}
\widetilde{\psi}(\{p\}) &= \frac{\langle \{p\}|\psi\rangle}{\sqrt{n!}\prod_i \sqrt{2E_{\bm{p}_i}}}, \\
\psi(\{x\})                   &= \frac{1}{\sqrt{n!}} \langle \{x\}|\psi\rangle
\end{aligned}
\end{equation}
and
\[
d\widetilde{\bm{p}}_i\equiv \frac{d\bm{p}_i}{(2\pi)^3\sqrt{2E_{\bm{p}_i}}}.
\]
The wave-functions $\widetilde{\psi}(\{p\})$ and $\psi(\{x\})$ are Fourier transforms of each other
\begin{equation}
\begin{aligned}
\psi(\{x\})                      & = \int \left(\prod_{i=1}^n  d\widetilde{\bm{p}}_i u_{m_i}(\bm{p}_i,s_i) e^{i\bm{p}_i\bm{x}_i}  \right) \widetilde{\psi}(\{p\}),\\
\widetilde{\psi}(\{p\}) & = \int \left(\prod_{i=1}^n d\bm{x}_i \frac{u^\dagger_{m_i}(\bm{p}_i,s_i)}{\sqrt{2E_{\bm{p}_i}}} e^{-i\bm{p}_i\bm{x}_i}\right)\psi(\{x\}). 
\end{aligned}
\label{eq:psi_Fourier_ndim}	                        
\end{equation}  
Imposing $\langle\psi|\psi\rangle=1$ the wave-functions $\widetilde{\psi}(\{p\})$ and $\psi(\{x\})$ are normalized as
\begin{equation}
\int \left(\prod_{i=1}^n\frac{d\bm{p}_i}{(2\pi)^3 } \right)|\widetilde{\psi}(\{p\})|^2=\int  \left(\prod_{i=1}^n d\bm{x}_i \right)|\psi(\{x\})|^2=1.
\end{equation}
Both $\widetilde{\psi}(\{p\})$ and $\psi(\{x\})$ are  anti-symmetric under an odd number of particle interchanges.

\subsection{The wave function of a nucleus}
\label{app:wave_function_nucleus}
The Fock state $|P_n\rangle$ of a nucleus, with four-momentum $P_n$ being in the $n$-th quantum state, can be written as a superposition of free nucleon states using their bound state wave-function  in the momentum representation $\widetilde{\psi'}_n$
\begin{equation}
|P_n\rangle=\int \left(\prod_{i=1}^{A} d\widetilde{\bm{p}}_i\right)\frac{\widetilde{\psi'}_n(\{p\})}{\sqrt{A!}}|\{p\}\rangle.
\label{eq:nucleus_state_1}
\end{equation}

The wave-function $\widetilde{\psi'}_n(\{p\})$ describes both the internal structure of the nucleus and its movement as a whole with three-momentum $\bm{p}=\sum_{i=1}^A\bm{p}_i$ and spin projection $s$.
Since the quantum state of $A$ interacting nucleons cannot depend on the motion of their center-of-mass, the wave-function $\widetilde{\psi'}_n(\{p\})$ can be factorized into a product of the wave-function $\widetilde{\psi}_n(\{p{^\star}\})$, describing the internal structure of the nucleus in its center-of-mass (the corresponding momenta are labeled by the upper index $\star$), and the wave-function $\Phi(p)$, describing the motion of the nucleus with momentum $\bm{p}$ and spin projection $s$, both encoded in the argument $p$ of $\Phi$
\begin{equation}
\widetilde{\psi}'_n(\{p\}) = \widetilde{\psi}_n(\{p{^\star}\})\Phi_n(p).
\label{eq:exclude_last_momentum}
\end{equation}
The factorization in~\cref{eq:exclude_last_momentum} makes sense for $A>1$.

The three-momentum of the $i$-th nucleon in the center-of-mass frame is given by $\bm{p}_i^\star$.
The $i$-th nucleon's momentum $\bm{p}_i$ in the laboratory system is given by 
\begin{equation}
\bm{p}_i = \bm{p}_i^\star+\bm{p}/A.
\label{eq:momentum_nucleon_lab}
\end{equation} 
The state in~\cref{eq:nucleus_state_1} can now be rewritten  as
\begin{equation}
|P_n\rangle  = \int \left(\prod_{i=1}^{A} d\widetilde{\bm{p}}_i^\star\right) \frac{\psi_n(\{p{^\star}\}) }{\sqrt{A!}}\Phi_n(p)|\{p\}\rangle.
\label{eq:nucleus_state_2}
\end{equation}
We take the wave-function $\Phi(p)$ of the form $$\Phi_n(p)=(2\pi)^3\sqrt{2P_n^0}\delta^3(\bm{p}-\bm{P}),$$ which corresponds to a nucleus with a definite momentum $\bm{P}$ and energy $P^0_n=E_{\bm{p}}+\varepsilon_n$, including excitation energy $\varepsilon_n$.
Then, the state in~\cref{eq:nucleus_state_2} is normalized similarly to~\cref{eq:1particle_state_norm} 
\begin{equation}
\langle P'_m|P_n\rangle = (2\pi)^3 2P^0_n \delta^3(\bm{P}-\bm{P}')\delta_{nm}
\label{eq:nucleus_state_norm_1}
\end{equation}
if the following normalization of the internal nucleus state $|n\rangle$ is adopted
\begin{equation}
\begin{aligned}
&\langle m|n\rangle   =\delta_{nm}\\
& = \int \left(\prod_{i=1}^{A}\frac{d\bm{p}_i^\star}{(2\pi)^3}\right) \widetilde{\psi}_n(\{p^\star\}) \widetilde{\psi}_m^*(\{p^\star\})(2\pi)^3\delta^3\left(\sum_{i=1}^A\bm{p}_i^\star\right).
\label{eq:nucleus_state_norm_2}
\end{aligned}
\end{equation}
The delta-function $\delta^3\left(\sum_{i=1}^A\bm{p}_i^\star\right)$ reduces the number of independent momenta in~\cref{eq:nucleus_state_norm_2} by one.

The states $|n\rangle$ and $|P_n\rangle$ describe the same realm at $\bm{P}=0$ yet still differ by normalization.
We define the former as
\begin{equation}
|n\rangle = \int \left(\prod_{i=1}^{A} d\widetilde{\bm{p}}_i^\star\right) \frac{\psi_n(\{p{^\star}\}) }{\sqrt{A!}} \Bigg[(2\pi)^3\delta^3\left(\sum_{i=1}^A\bm{p}_i^\star\right)\Bigg]^{1/2}|\{p^\star\}\rangle
\label{eq:n_state}
\end{equation}
which agrees with the normalization in~\cref{eq:nucleus_state_norm_2}. 
\section{Derivation of the $\ensuremath{\nu A\to \nu A}$ cross-section}
\label{app:cross_section}
\subsection{Scattering amplitude}
\label{app:amplitude}
An effective SM Lagrangian  in the four-fermion approximation 
\begin{equation}
\mathcal{L}(x) = \frac{G_F}{\sqrt{2}} L_\mu(x)H^\mu(x)
\label{eq:lepton_current}
\end{equation}
should be accurate enough for the scattering of a low energy neutrino  off of a nucleus. 
In~\eqref{eq:lepton_current} 
\[
L_\mu(x)= \; \normord{\overline{\psi}_\nu(x)\gamma_\mu(1-\gamma_5)\psi_\nu(x)}
\] 
and 
\begin{equation}
H^\mu(x) = \sum_{f=n,p}\normord{ \overline{\psi}_f(x)\gamma^\mu\left(g_V^f-g_A^f\gamma_5\right)\psi_f(x)}
\label{eq:hadronic_field_current}
\end{equation}
are weak currents of neutrino and  nucleons, respectively, written in the normal ordering represented by colons.
The quantum fields $\psi_\nu(x)$ and $\psi_{n,p}(x)$  correspond to the neutrino and nucleons, respectively.

The $\mathbb{S}$-matrix amplitude $\langle P'_m,k'|\mathbb{S}|P_n,k\rangle$, to the first order of $G_F$, reads
\begin{equation}
\begin{aligned}
\langle P'_m,k'|\mathbb{S}|P_n,k\rangle &= (2\pi)^4\delta^4(q+P_n-P'_m)i\mathcal{M}_{mn},\\
i\mathcal{M}_{mn} & = i\frac{G_F}{\sqrt{2}}l_\mu H^\mu_{mn},
\label{eq:matrix_element_1}
\end{aligned}
\end{equation}
where
\begin{equation}
l_\mu(k,k') = \overline{u}(k',-1)\gamma_\mu(1-\gamma_5)u(k,-1)
\label{eq:leptonic_current}
\end{equation}
and 
\begin{equation}
H^\mu_{mn}(P_n,P'_m) = \langle P'_m|H^\mu(0)|P_n\rangle.
\label{eq:hadronic_current_1}
\end{equation}
Using \eqref{eq:1particle_state_norm}, \eqref{eq:hadronic_field_current} and the anti-symmetric nature of the wave-function, the hadronic current in \eqref{eq:hadronic_current_1} can be found 
\begin{equation}
H^\mu_{mn}(P,P') = 2\sqrt{P^0_n P^{'0}_m }h^\mu_{mn}
\end{equation}
with 
\begin{equation}
\begin{aligned}
h^\mu_{mn} = &  \sum_{k=1}^{A} \int \left(\prod_{j=1}^{A}\frac{ d\bm{p}_j^\star }{(2\pi)^3}\right)
						 \frac{\overline{u}(\bm{p}_{k}+\bm{q},s_{k})O^\mu_k u(\bm{p}_{k},r_{k})}{\sqrt{2E_{\bm{p}_{k}}  2E_{\bm{p}_{k}+\bm{q} } }}\\
						& \times (2\pi)^3\delta^3(\sum_{l=1}^A\bm{p}_l^\star)\widetilde{\psi}_m^*(\{p^{(k)}_\star\} ) \widetilde{\psi}_n(\{p_\star\}),
\end{aligned}
\label{eq:hmunu_1}
\end{equation}
where 
\begin{equation}
\begin{aligned}
O^\mu_k & =\gamma^\mu\left(g_V^k-g_A^k\gamma_5\right)\\
              &=\gamma^\mu\left(g_L^k(1-\gamma_5)+g_R^k(1+\gamma_5)\right)
\end{aligned}
\end{equation}
and the couplings $g_{V,A}^k$ are equal to $g_{V,A}^{p/n}$ when the index $k$ points to a proton/neutron. 
Left- and right-chirality couplings are expressed via vector $g_V^{p/n}$ and axial $g_A^{p/n}$ couplings as
\begin{equation}
\begin{aligned}
g_L^{p/n} & =\frac{1}{2}\left(g_V^{p/n}+g_A^{p/n}\right),\\
g_R^{p/n} & =\frac{1}{2}\left(g_V^{p/n}-g_A^{p/n}\right).
\end{aligned}
\label{eq:left_right_couplings}
\end{equation}
In the SM these couplings read
\begin{equation}
\begin{aligned}
g_V^p  & =   \frac{1}{2}-2\sin^2\theta_W,      					   & g_A^p &=  \frac{1}{2},  \\
g_L^p  & =    \frac{1}{2}\left(1-2\sin^2\theta_W\right),  & g_R^p &=   -\sin^2\theta_W,\\ 
g_V^n & =  -\frac{1}{2},                                                           & g_A^n &=-\frac{1}{2},\\
g_L^n & = -1,                                                                             & g_R^n &= 0.
\end{aligned}
\label{eq:sm_couplings}
\end{equation}
The arguments  of   $\widetilde{\psi}_m^*(\{p^{(k)}_\star\} )$ and $\widetilde{\psi}_n(\{p_\star\})$ are  $n$-tuples defined as $\{p_\star\}=(p_1^\star\dots p_A^\star)$, where its $i$-th element, $p_i^\star=(\bm{p}_i^\star,r_i)$ and $\{p^{(k)}_\star\}$, is identical to $\{p_\star\}$ except for its $k$-th element, which reads as $(\bm{p}_{k}^\star+\bm{q},s_k)$.
The three momentum $\bm{p}_{k}$, used in the argument of the Dirac spinor $u$, is the $k$-th nucleon momentum in the laboratory frame given by~\eqref{eq:momentum_nucleon_lab}.

The hadronic current, corresponding to neutrino-nucleus scattering, is a sum of currents $\overline{u}(\bm{p}_k+\bm{q},s_k)O^\mu_k u(\bm{p}_k,r_k)$ corresponding to the scattering of a neutrino off of the $k$-th nucleon with momentum in the laboratory $\bm{p}_k$ and spin projection $r_k$.
The probability amplitude to find a nucleon in the $|P_n\rangle$ state of the nucleus with these quantum numbers is just the wave-function $\widetilde{\psi}_n(\{p_\star\})$ in the momentum representation, which depends on momenta in the nucleus center-of-mass frame.

The outgoing nucleon has a three-momentum in the laboratory of $\bm{p}_k+\bm{q}$, and, in general, an arbitrary spin projection $s_k$.
The corresponding probability amplitude to find a nucleon with exactly these quantum numbers is given similarly by the wave-function $\widetilde{\psi}_m^*(\{p^{(k)}_\star\})$.

The denominator  $\sqrt{2E_{\bm{p}_k}  2E_{\bm{p}_k+\bm{q}}}$ depends on the energies of the initial and final nucleons in the laboratory frame, and automatically accounts for the normalization of Dirac spinors $u^\dagger(\bm{p},s)u(\bm{p},s)=2E_{\bm{p}}$.

The equal momenta of the initial and final state spectator nucleons are integrated out with the weight given by a product of initial and final state wave-functions.

To proceed further, let us make the following simplifications. 
The current $\overline{u}(\bm{p}_k+\bm{q},s_k)O^\mu_k u(\bm{p}_k,r_k)$ could be factorized out from the integral  at an effective momentum $\bm{p}_k$ which we approximate  to be given by a solution of~\cref{eq:energy_conservation_nucleon_nucleus2}.
Also, we assume that the spin and momenta structures of $\widetilde{\psi}_n$ could be factorized into a product $\widetilde{\psi}_n$ and $\chi_n$
\begin{equation}
\label{eq:factorize_spin}
\widetilde{\psi}_n(\{p_\star\})=\widetilde{\psi}_n(\{\bm{p}_\star\})\chi_n(\{r\}),
\end{equation}
which are functions  of two $n$-tuples $\{\bm{p}_\star\}=(\bm{p}_1^\star\dots{} \bm{p}_A^\star)$ and $\{r\}=(r_1\dots r_A)$, respectively.
The spin-functions can be normalized as follows
\begin{equation}
\label{eq:spin_functions_norm}
\chi_m^\dagger(\{r\})\chi_n(\{r\}) = \delta_{nm}.
\end{equation}
Thus,~\eqref{eq:hmunu_1} can be rewritten as
\begin{equation}
\begin{aligned}
&h^\mu_{mn} =   \sum_{k=1}^{A} \frac{\overline{u}(\bm{p}+\bm{q},s_{k})O^\mu_k u(\bm{p},r_{k})}{\sqrt{2E_{\bm{p}}  2E_{\bm{p}+\bm{q} } }}\chi^*_m(\{r^{(k)}\})\chi_n(\{r\})\\
                         & \times\int \left(\prod_{j=1}^{A}\frac{ d\bm{p}_j }{(2\pi)^3}\right) (2\pi)^3\delta^3(\sum_{l=1}^A\bm{p}_l) 
                         \widetilde{\psi}_m^*(\{\bm{p}^{(k)}_\star\} ) \widetilde{\psi}_n(\{\bm{p}_\star\}),\\
\end{aligned}
\label{eq:hmunu_2}
\end{equation}
where $\{r^{(k)}\}$ is an $n$-tuple identical to $\{r\}$, except its $k$-th element, which is equal to $s_k$.

A further insight could be gained by observing that one can rewrite the multidimensional integral  in~\eqref{eq:hmunu_2} as the matrix element 
\begin{equation}
\begin{aligned}
&\int \left(\prod_{j=1}^{A}\frac{ d\bm{p}_j^\star }{(2\pi)^3}\right) (2\pi)^3\delta^3(\sum_{l=1}^A\bm{p}_l^\star)\widetilde{\psi}_m^*(\{\bm{p}^{(k)}_\star\} ) \widetilde{\psi}_n(\{\bm{p}_\star\})\\
&=\langle m|e^{i\bm{q}\hat{\bm{X}}_{k}}|n\rangle\equiv f^k_{mn}(\bm{q}),
\end{aligned}
\label{eq:hmunu_3}
\end{equation}
where $\hat{\bm{X}}$ is the three-coordinate operator of the $k$-th nucleon.

\cref{eq:hmunu_3} provides a clue in understanding the appearance of coherent and incoherent regimes in neutrino-nucleus elastic and inelastic scattering.
 
A derivation of~\cref{eq:hmunu_3}  is facilitated if the following equality is observed
\begin{equation}
\langle\bm{p}|e^{i\bm{q}\hat{\bm{X}}} = \frac{\sqrt{2E_{\bm{p}}}} {\sqrt{2E_{\bm{p}+\bm{q} }}}\langle\bm{p}+\bm{q}|.
\label{eq:shift_q}
\end{equation}
\cref{eq:shift_q} can be proven with help of~\cref{eq:x_definition_nonrel}.
Using~\cref{eq:shift_q,eq:n_state} the matrix element $\langle m|e^{i\bm{q}\hat{\bm{X}}_{k}}|n\rangle$ can be calculated. 

Therefore, combining~\cref{eq:matrix_element_1,eq:hadronic_current_1,eq:hmunu_1,eq:hmunu_3} one gets the matrix element of elastic neutrino-nucleus scattering 
\begin{equation}
\begin{aligned}
 i\mathcal{M}_{mn} & = i\frac{G_F}{\sqrt{2}} \left(\frac{P^0_n }{E_{\bm{p}}}\frac{ P^{'0}_m }{E_{\bm{p}+\bm{q} } }\right)^{1/2} \! l_\mu(k,k')\sum_{k=1}^{A}\langle m|e^{i\bm{q}\hat{\bm{X}}_{k}}|n\rangle\\
 &\times\chi^*_m(\{r^{(k)}\})\chi_n(\{r\}) \overline{u}(\bm{p}+\bm{q},s_{k})O^\mu_k u(\bm{p},r_{k}).
\end{aligned}
\label{eq:matrix_element_2}
\end{equation}
We introduce the following notation for economy of space:
\begin{equation}
 \chi^*_m(\{r^{(k)}\})\chi_n(\{r\})\equiv \lambda^{mn}(s,r).
\label{eq:lambda_def}
\end{equation}
In general, the scattered nucleus may have a final spin state different with respect to the initial. 
We assume in what follows that  initial and final states of the nucleus are eigenstates of the spin operator with quantum numbers $(J,J_3)$.
One might observe that if $m=n$, then the amplitude $\lambda^{mn}(s,r)=\delta_{sr}$ for appropriate normalization of the spin wave-function (see the normalization used in~\cref{eq:spin_functions_norm}).
We denote for $m\ne n$ the corresponding amplitude as $\lambda^{mn}_{sr}$.
Therefore, for any $m,n$
\begin{equation}
\lambda^{mn}(s,r)= \delta_{mn}\delta_{sr}+(1-\delta_{mn})\lambda^{mn}_{sr}.
\label{eq:spin_functions_norm2}
\end{equation}
The multiplier in~\cref{eq:matrix_element_2} can be rewritten, factoring out the leading order term $m_A/m_N$ and the factor $C_{mn,1}$ of the order of unity defined as
\begin{equation}
C_{1,mn}^{1/2} = \left(\frac{P^0_n }{E_{\bm{p}}}\frac{ P^{'0}_m }{E_{\bm{p}+\bm{q} } }\right)^{1/2}\frac{m_N}{m_A}
\label{eq:C1_def}
\end{equation}
Using~\cref{eq:hmunu_3,eq:lambda_def,eq:C1_def}, one can represent~\cref{eq:matrix_element_2} as in~\cref{eq:matrix_element}. 
\subsection{Cross-sections}
\label{app:cross-section}
The cross-section corresponding to the matrix element in~\cref{eq:matrix_element_1} reads
\begin{equation}
\begin{aligned}
&\frac{d^2\sigma_{mn}}{dE_\nu^\prime d\cos\theta} \\
&=\frac{-E^\prime_\nu|i\mathcal{M}_{mn}|^2}{2^5\pi E_\nu (m_A+\varepsilon_n)}\frac{ \delta(E_\nu-E^\prime_\nu-T_A-\Delta\varepsilon_{mn}) }{m_A+T_A+\varepsilon_m},
\end{aligned}
\end{equation} 
where  all kinematic variables are given in the laboratory frame in which the initial nucleus is assumed to be at rest, $E^\prime_\nu$ is given by~\cref{eq:outgoing_neutrino_energy_lab} and $\Delta\varepsilon_{mn}=\varepsilon_m-\varepsilon_n$.
The kinetic energy $T_A$ of the scattered nucleus is given by~\cref{eq:kinetic_energy_exact,eq:kinetic_energy}.
Integration over $E_\nu^\prime$ can be done with help of a Dirac $\delta$-function, providing energy conservation, thus yielding
\begin{equation}
\label{eq:cross_section_costheta}
\begin{aligned}
\frac{d\sigma_{mn}}{d\cos\theta} &= \frac{-|i\mathcal{M}_{mn}|^2}{2^5\pi (m_A+\varepsilon_n)}\frac{E^\prime_\nu (m_A+T_A)}{E_\nu (m_A+T_A+\varepsilon_m)}\\
&\times\frac{1}{m_A+E_\nu(1-\cos\theta)-\Delta\varepsilon_{mn}},
\end{aligned}
\end{equation} 
One can obtain $d\sigma_{mn}/dT_A$ using a very accurate approximation given in~\cref{eq:kinetic_energy}
\begin{equation}
\begin{aligned}
\frac{d\sigma_{mn}}{dT_A} & = \frac{d\sigma_{mn}}{d\cos\theta}\frac{d\cos\theta}{dT_A}
                                                   =-\frac{d\sigma_{mn}}{d\cos\theta}\frac{m_A}{E_\nu(E_\nu-\Delta\varepsilon_{mn})}\\
                                                   &= \frac{|i\mathcal{M}_{mn}|^2}{2^5\pi E_\nu^2m_A} C_{2,mn},
\end{aligned}
\label{eq:cross_section_TA}
\end{equation}
where 
\begin{equation}
\label{eq:C2_mn}
\begin{aligned}
C_{2,mn} &= \frac{E_\nu'}{E_\nu-\Delta\varepsilon_{mn}}\frac{\left(1+\frac{T_A}{m_A}\right)\left(1+\frac{\varepsilon_n}{m_A}\right)^{-1}}{ \left(1+\frac{T_A+\varepsilon_m}{m_A}\right)}\\
                  &\times \left(1+\frac{E_\nu(1-\cos\theta)-\Delta\varepsilon_{mn}}{m_A}\right)^{-1}
\end{aligned}
\end{equation}
is of the order of unity.

Combining~\cref{eq:matrix_element,eq:cross_section_TA,eq:hmunu_3} one gets an observable differential cross-section defined in~\cref{eq:cross-section_2}
\begin{equation}
\label{eq:cross_section_2}
\begin{aligned}
&\frac{d\sigma}{dT_A}  = \frac{G_F^2 m_A}{2^6\pi m_N^2E_\nu^2} \\
\times&\sum_{k,j=1}^{A} \sum_{n}\omega_nC_{1,mn}C_{2,mn}\left(  f_{nn}^k f_{nn}^{j*}\sum_{r} (l,h^k_{rr}) \sum_{s} (l,h^j_{ss})^\dagger\right.\\
&\left. +\sum_{m\ne n}  f_{mn}^k f_{mn}^{j*}\sum_{sr}\lambda^{mn}_{sr}(l,h^k_{sr})\left(\sum_{s'r'}\lambda^{mn}_{s'r'}(l,h^j_{s'r'})\right)^\dagger
\right)
\end{aligned}
\end{equation}
expressed through the scalar products $(l,h^{p/n}_{sr})$ of 4-vectors with components $l^\mu(k,k')$ given by~\cref{eq:leptonic_current} and 
\begin{equation}
\label{eq:nucleon_current}
(h^{p/n}_{sr})_\mu = \overline{u}({\bm{p}}+\bm{q},s)O_\mu^{p/n} u({\bm{p}},r)
\end{equation}
where  $\bm{p}$ is a solution of~\cref{eq:energy_conservation_nucleon_nucleus2}.
In~\cref{eq:nucleon_current} a superscript $p$ or $n$ appears when the index $k$ in $h^k_{sr}$ from~\cref{eq:cross_section_2} points to a proton or to a neutron, respectively.
When an index $k$ or $j$ in~\cref{eq:cross_section_2} points to a proton/neutron, the form-factors $f_{mn}^k$ should be read as $f^{p/n}_{mn}$, correspondingly.

Each of the $|(l,h^{p/n})|^2$ terms given by~\cref{eq:scalar_products_helicity,eq:scalar_products_sigma3} yields the common factor  $64(s-m_N^2)^2$, where $s=(p+k)^2$ is the total energy squared in the neutrino-nucleon center-of-mass frame, and $m_N$ is the mass of the nucleon.
In the leading non-relativistic approximation this factor can be approximated as $2^8 m_N^2E_\nu^2$.
We denote a correction to this formula by a factor $C_{3,mn}$, accounting for the fact that the nucleon in the initial state has a non-zero three-momentum
\begin{equation}
(s-m_N^2)^2 = 4m_N^2E_\nu^2C_{3,mn}.
\end{equation}
In what follows we denote by $g^{mn}$ the product of correction factors 
\begin{equation}
g^{mn} = C_{1,mn} C_{2,mn} C_{3,mn}
\label{eq:g_functions}
\end{equation}
which is of the order of unity.

Following our discussion of~\cref{eq:matrix_element} we identify the second and third lines of~\cref{eq:cross_section_2} as contributing to the coherent and incoherent cross-sections.
The factor $g^{mn}$ is, in general, different for coherent and incoherent terms.
We take out these factors from the double summation at their effective values denoted by $g_c$ and $g_i$ for coherent and incoherent terms, respectively.

The summation over $n$ in the second line of~\cref{eq:cross_section_2} leads to the form-factors averaged over all initial states
\begin{equation}
\label{eq:form_factors_1}
\sum_n \omega_n f_{nn}^k f_{nn}^{j*}=
\left\{
\begin{matrix}
|F_{p/n}(\bm{q})|^2,                                 & (k,j)\to pp\text{ or }nn,\\
F_{p}(\bm{q})F_{n}^*(\bm{q}),            & (k,j)\to pn,\\
F_{n}(\bm{q})F_{p}^*(\bm{q}),            & (k,j)\to np.\\
\end{matrix}
\right.
\end{equation}
Therefore, the second line of~\cref{eq:cross_section_2} can be re-written as
\begin{equation}
\label{eq:cross_section_coherent_term}
\left| \sum_{k=1}^Z\sum_r (l, h^p_{rr})F_p + \sum_{k=1}^N \sum_{r}(l, h^n_{rr})F_n\right|^2.
\end{equation}

Let us work out the incoherent scattering encoded in the third line of~\cref{eq:cross_section_2}.
A summation over $m,n$  cannot be done without a model for $\lambda^{mn}_{sr}$.
If $\lambda^{mn}_{sr}$ would not depend on $m,n$ the corresponding summation could be performed as follows.

Consider the case when $k$ and $j$ point to the same type of the nucleon, for example, to a proton. 

If $k=j$, then
\begin{equation}
\label{eq:cross_section_incoherent_term1}
\begin{aligned}
\sum_n &\omega_n\sum_{m\ne n} f_{mn}^k f_{mn}^{k*} =\sum_n \omega_n \left[\sum_{m} f_{mn}^k f_{mn}^{k*} -f_{nn}^k f_{nn}^{k*}\right]\\
&=\sum_n \omega_n \left[ \langle n|e^{-i\bm{q}\bm{X}_k}\sum_{m}|m\rangle\langle m|e^{i\bm{q}\bm{X}_k}|n\rangle\right] -|F_{p}(\bm{q})|^2\\
&=1-|F_{p}(\bm{q})|^2,
\end{aligned}
\end{equation}
accounting for the equality $\sum_m|m\rangle \langle m|=\hat{I}$, using~\cref{eq:form_factors_1}  and normalizations  in~\cref{eq:nucleus_state_norm_2} and $\sum_n \omega_n=1$.

If $k\ne j$ then following a consideration similar to~\cref{eq:cross_section_incoherent_term1} one may find that 
 \begin{equation}
 \label{eq:cross_section_incoherent_term2}
 \sum_n \omega_n\sum_{m\ne n} f_{mn}^k f_{mn}^{j*} = \langle\text{cov}(e^{-i\bm{q}\hat{\bm{X}}_j},e^{i\bm{q}\hat{\bm{X}}_k})\rangle_p
 \end{equation}
 where the right-hand-side of~\cref{eq:cross_section_incoherent_term2} is a covariance of  quantum operators $e^{-i\bm{q}\hat{\bm{X}}_j}$ and $e^{i\bm{q}\hat{\bm{X}}_k}$ on $|n\rangle$, whose state reads
 \begin{equation}	
 \begin{aligned}
&\text{cov}_{nn}(e^{-i\bm{q}\hat{\bm{X}}_j},e^{i\bm{q}\hat{\bm{X}}_k}) \\
&=\langle n|e^{-i\bm{q}\hat{\bm{X}_j}}e^{i\bm{q}\hat{\bm{X}}_k}|n\rangle - \langle n|e^{i\bm{q}\hat{\bm{X}}_k}|n\rangle \langle n|e^{-i\bm{q}\hat{\bm{X}}_j}|n\rangle.
\end{aligned}
\end{equation}
The subscript $p$ in~\cref{eq:cross_section_incoherent_term2} refers to a proton.

The averaging $\langle\dots\rangle$ in~\cref{eq:cross_section_incoherent_term2} is given by
\begin{equation}
\langle\text{cov}(e^{-i\bm{q}\hat{\bm{X}}_j},e^{i\bm{q}\hat{\bm{X}}_k})\rangle_p = \sum_n \omega_n \text{cov}_{nn}(e^{-i\bm{q}\hat{\bm{X}}_j},e^{i\bm{q}\hat{\bm{X}}_k}).
\end{equation}
 At both, $\bm{q}\to 0$ and $\bm{q}\to\infty$
 \begin{equation}
 \begin{aligned}
 &\lim_{\bm{q}\to 0}\langle\text{cov}(e^{-i\bm{q}\hat{\bm{X}}_j},e^{i\bm{q}\hat{\bm{X}}_k})\rangle_p = 0,\\
 & \lim_{\bm{q}\to \infty}\langle\text{cov}(e^{-i\bm{q}\hat{\bm{X}}_j},e^{i\bm{q}\hat{\bm{X}}_k})\rangle_p=0.
 \end{aligned} 
 \label{eq:cross_section_incoherent_term3}
 \end{equation}
 In the case of weak correlations of nucleons in a nucleus, the covariances, like in~\cref{eq:cross_section_incoherent_term2} vanish.
 For example, in models like the nuclear shell model, where a multi-particle wave-function is constructed in terms of a product of one-particle wave-functions, the covariance in~\cref{eq:cross_section_incoherent_term2} is identically zero.

 The smallness of the covariance in~\cref{eq:cross_section_incoherent_term2} is the reason why the inelastic cross-section is, to good accuracy, linearly dependent on the number of nucleons.
 In what follows the covariance terms are neglected.

 The same considerations apply to the  scattering on a neutron.
 It is straightforward to show that in the case of mixing neutron and proton amplitudes one gets (let $k$ point to a proton and $j$ point to a neutron, and now automatically $k\ne j$)
 \begin{equation}
 \label{eq:cross_section_incoherent_term4}
 \begin{aligned}
 \sum_n \omega_n\sum_{m\ne n} f_{mn}^k f_{mn}^{j*} =\langle\text{cov}(e^{i\bm{q}\hat{\bm{X}}_k},e^{-i\bm{q}\hat{\bm{X}}_j})\rangle_{pn}\\
 \end{aligned}
 \end{equation}
 which can also be neglected.

 As mentioned above the exact summation should consider the spin amplitude $\lambda^{mn}_{sr}$.
 We approximate the summation by replacing $\lambda^{mn}_{sr}$ by its average value $\lambda^{p/n}_{sr}$ for protons and neutrons, respectively.
  
Therefore,  the third line of~\cref{eq:cross_section_2} reads
\begin{equation}
\label{eq:cross_section_incoherent_term5}
\begin{aligned}
 &\sum_{k=1}^Z \sum_{sr}|\lambda^p_{sr}|^2|(l,h^p_{sr})|^2(1-|F_p|^2)\\
+&\sum_{k=1}^N \sum_{sr}|\lambda^n_{sr}|^2|(l, h^n_{sr})|^2(1-|F_n|^2).
\end{aligned}
\end{equation}
Combining~\cref{eq:cross_section_2,eq:cross_section_coherent_term,eq:cross_section_incoherent_term5}, one gets the differential cross-section in~\cref{eq:cross-section_3}.

\section{Calculation of the scalar product $(l,h)$}
\label{app:matrix_element_calculation}
The third line of~\cref{eq:cross-section_3} prompts us to calculate the scalar product of two currents  $\overline{u}(k')O^\mu u(k)\cdot \overline{u}(p')O'_\mu u(p)$, where $O_\mu, O'_\mu$ are  Dirac matrices.
The use of  a standard powerful technique, which consists of the calculation of traces of Dirac $\gamma$-matrices, is not helpful for this problem.
This is because  all four momenta $k,k',p$ and $p'$ are different and one cannot use the well-known formula for Dirac spinors  
\begin{equation*}
u(\bm{p},r)\overline{u}(\bm{p},r) =\frac{1}{2}\left(\slashed{p}+m\right)\left(1+\gamma_5\slashed{s}_r\right),
\end{equation*} 
where $s_r$ is four-vector of the fermion spin.

To simplify intermediate formulas,  we calculate the scalar product of the neutrino and nucleon currents  in their center-of-mass frame, where energies of incoming and outgoing fermions are equal.
In what follows in this section all quantities depending on kinematic variables are given in the neutrino-nucleon center-of-mass frame. 

Energies $E_\nu$ and $E_N$, of the neutrino  and nucleon, respectively, read
\begin{equation}
\begin{aligned}
E_\nu & = \frac{s-m^2}{2\sqrt{s}}, &E_N & = \frac{s+m^2}{2\sqrt{s}},  
\end{aligned}
\end{equation}
where $s = (p+k)^2$ and $m$ gives the mass of the nucleon.

In the Dirac basis the spinor of a nucleon with three-momentum $\bm{p}$ and index $r=\pm 1$ reads
\begin{equation}
	\label{eq:nucleon_spinor_center_of_mass}
	u(\bm{p},r) =
	\begin{pmatrix}
		\lambda_+\\
		\lambda_-\alpha_{\bm{p}}
	\end{pmatrix}
	\chi_r(\bm{p}),
\end{equation}
where $\lambda_\pm=\sqrt{E_N\pm m}$ and $\alpha_{\bm{p}}=\bm{n}_{\bm{p}}\cdot\bm{\sigma}$, in which  $\bm{n}_{\bm{p}}$ is a unit vector along $\bm{p}$, and $\bm{\sigma}=(\sigma_1,\sigma_2,\sigma_3)$ is a three-vector of Pauli matrices.
The index $r$ enumerates two linearly independent two-spinors $\chi_r(\bm{p})$.

The vector and axial currents of the nucleon read
\begin{equation}
\label{eq:nucleon_vector_axial_currents1}
\begin{aligned}
\overline{u}(\bm{p}', r')\gamma^\mu u(\bm{p}, r)  & \equiv V^\mu_{r'r}(\bm{p}',\bm{p}),\\
\overline{u}(\bm{p}', r')\gamma^\mu\gamma^5 u(\bm{p}, r)  & \equiv A^\mu_{r'r}(\bm{p}',\bm{p}),\\
\end{aligned}
\end{equation}
where 
\begin{equation}
\label{eq:nucleon_vector_axial_currents2}
\begin{aligned}
V^0_{r'r}(\bm{p}',\bm{p}) & = \chi^\dagger_{r'}(\bm{n}_{\bm{p}'})\left[\lambda_+^2+\lambda_-^2\alpha_{\bm{p}'}\alpha_{\bm{p}}\right]\chi_{r}(\bm{n}_{\bm{p}}),\\
\bm{V}_{r'r}(\bm{p}',\bm{p}) & = \chi^\dagger_{r'}(\bm{n}_{\bm{p}'})\lambda_+\lambda_-\left[\bm{\sigma}\alpha_{\bm{p}}+\alpha_{\bm{p}'}\bm{\sigma}	\right]\chi_{r}(\bm{n}_{\bm{p}}),\\
A^0_{r'r}(\bm{p}',\bm{p}) & = \chi^\dagger_{r'}(\bm{n}_{\bm{p}'})\lambda_+\lambda_-\left[\alpha_{\bm{p}'}+\alpha_{\bm{p}}\right]\chi_{r}(\bm{n}_{\bm{p}}),\\
\bm{A}_{r'r}(\bm{p}',\bm{p}) & = \chi^\dagger_{r'}(\bm{n}_{\bm{p}'})\left[\lambda_+^2\bm{\sigma}+\lambda_-^2\alpha_{\bm{p}'}\bm{\sigma}\alpha_{\bm{p}}\right]\chi_{r}(\bm{n}_{\bm{p}}).\\
\end{aligned}
\end{equation}
The neutrino spinor, vector, and axial currents read analogously to~\cref{eq:nucleon_spinor_center_of_mass,eq:nucleon_vector_axial_currents1,eq:nucleon_vector_axial_currents2} with the replacement $\lambda_\pm\to \sqrt{E_\nu}$.

Unit vectors along three-momenta of the incoming and outgoing neutrino and nucleon are defined as
\begin{equation}
\begin{aligned}
\bm{n}_{\bm{k}} & =(0,0,1), &&\bm{n}_{\bm{p}} & =-\bm{n}_{\bm{k}},\\
\bm{n}_{\bm{k}'} & = (\cos\varphi\sin\theta,\sin\varphi\sin\theta,\cos\theta), &&\bm{n}_{\bm{p}'} & = -\bm{n}_{\bm{k}'}. 
\end{aligned}
\end{equation}
It is convenient to specify a basis of two-component spinors $\chi_r$ to perform the calculations in~\cref{eq:nucleon_vector_axial_currents1}. 
Summation over $r,r'$ in the incoherent term of~\cref{eq:cross-section_3} are simpler in the helicity basis in which $r,r'$ are helicity eigenvalues.
The coherent term of~\cref{eq:cross-section_3} requires consideration of the nucleon current with conservation of spin projection on the given axis.
For this purpose a basis of $\chi_r$ two-spinors, which are eigenstates of the $\sigma_3=\left(\bm{n}_{\bm{k}}\cdot\bm{\sigma}\right)$ matrix, is more appropriate.
It is apparent that the physical observable does not depend on the basis chosen.
\subsection{Helicity basis}
In the helicity basis the two-spinor $\chi_{r}(\bm{p})$ is an eigenvector of the helicity operator $\bm{n}_{\bm{p}}\cdot\bm{\sigma}$
\begin{equation}
\bm{n}_{\bm{p}}\cdot\bm{\sigma} \chi_{r}(\bm{n}_{\bm{p}}) = r\chi_{r}(\bm{n}_{\bm{p}})
\end{equation}
with an eigenvalue $r=\pm 1$, known as the helicity or doubled spin projection on a particle's three-momentum. 
Two-component normalized to unity spinors $\chi$ corresponding to the incoming and outgoing neutrino and nucleon with  definite  helicities in their center-of-mass frame, can be read  
\begin{equation}
\label{eq:helicity_basis}
\begin{aligned}
\chi_{+}(\bm{n}_{\bm{k}}) & = \begin{pmatrix} 1 \\ 0\end{pmatrix}, 
& \chi_{+}(\bm{n}_{\bm{k}'}) & = \begin{pmatrix} \cos\frac{\theta}{2}\\ e^{i\varphi}\sin\frac{\theta}{2}\end{pmatrix},\\
\chi_{-}(\bm{n}_{\bm{k}}) & = \begin{pmatrix} 0 \\ 1\end{pmatrix}, & \chi_{-}(\bm{n}_{\bm{k}'}) & = \begin{pmatrix} -e^{-i\varphi}\sin\frac{\theta}{2} \\ \cos\frac{\theta}{2}\end{pmatrix}, \\
\chi_{+}(\bm{n}_{\bm{p}}) & = \begin{pmatrix} 0 \\ -1\end{pmatrix}, & \chi_{+}(\bm{n}_{\bm{p}'}) & = \begin{pmatrix} \sin\frac{\theta}{2} \\ -e^{i\varphi}\cos\frac{\theta}{2}\end{pmatrix},\\ 
\chi_{-}(\bm{n}_{\bm{p}}) & = \begin{pmatrix} 1 \\ 0\end{pmatrix}, & \chi_{-}(\bm{n}_{\bm{p}'}) & = \begin{pmatrix} e^{-i\varphi}\cos\frac{\theta}{2} \\ \sin\frac{\theta}{2}\end{pmatrix}.
\end{aligned}
\end{equation}
Let us display explicitly the vector and axial currents in the helicity basis from~\cref{eq:nucleon_vector_axial_currents1,eq:nucleon_vector_axial_currents2} for the nucleon, denoting the basis by the $\chi$ superscript 
\begin{equation}
\label{eq:nucleon_vector_axial_currents3}
\begin{aligned}
V^\chi_{++} & = 2\left(c_{\theta/2}e^{-i\varphi}E_N,\left(-s_{\theta/2},is_{\theta/2},-c_{\theta/2}e^{-i\varphi}\right)E_\nu\right),\\
V^\chi_{--} & = 2\left(c_{\theta/2}e^{+i\varphi}E_N,\left(-s_{\theta/2},-is_{\theta/2},-c_{\theta/2}e^{+i\varphi}\right)E_\nu\right),\\
V^\chi_{+-} & = 2m\left(s_{\theta/2},0,0,0\right),\\
V^\chi_{-+} & = 2m\left(-s_{\theta/2},0,0,0\right),\\
A^\chi_{++} & = 2\left(c_{\theta/2}e^{-i\varphi}E_\nu,\left(-s_{\theta/2},is_{\theta/2},-c_{\theta/2}e^{-i\varphi}\right)E_N\right),\\
A^\chi_{--} & = 2\left(-c_{\theta/2}e^{+i\varphi}E_\nu,\left(s_{\theta/2},is_{\theta/2},c_{\theta/2}e^{+i\varphi}\right)E_N\right),\\
A^\chi_{+-} & = 2m\left(0,-c_{\theta/2}e^{-i\varphi},-ic_{\theta/2}e^{-i\varphi},s_{\theta/2}\right),\\
A^\chi_{-+} & = 2m\left(0,-c_{\theta/2}e^{+i\varphi},ic_{\theta/2}e^{+i\varphi},s_{\theta/2}\right)\\
\end{aligned}
\end{equation}
and neutrino
\begin{equation}
\label{eq:neutrino_vector_axial_currents}
\begin{aligned}
V^\chi_{--} & = 2E_\nu\left(c_{\theta/2},s_{\theta/2}e^{+i\varphi},-is_{\theta/2}e^{+i\varphi},c_{\theta/2}\right),\\
A^\chi_{--} & = 2E_\nu\left(-c_{\theta/2},-s_{\theta/2}e^{+i\varphi},is_{\theta/2}e^{+i\varphi},-c_{\theta/2}\right),\\
\end{aligned}
\end{equation}
where for the sake of compactness 
\begin{equation}
c_{\theta/2}\equiv \cos\frac{\theta}{2}, \quad s_{\theta/2}\equiv \sin\frac{\theta}{2}.
\end{equation}
For a neutrino, assuming its vanishing mass,
\begin{equation}
V_{+-}=V_{-+}=A_{+-}=A_{-+}=0
\end{equation}
manifesting neutrino helicity conservation in weak interactions. 
In~\cref{eq:neutrino_vector_axial_currents} only the left-handed neutrino currents required to calculate the elastic neutrino-nucleus cross-section are shown.

Now it is straightforward to calculate the scalar product $(l,h_{r'r})$ equal to
\[
\overline{u}(k',-1)\gamma^\mu (1-\gamma^5)u(k,-1)\cdot\overline{u}(p',r')\gamma_\mu (g_V-g_A\gamma^5)u(p,r)
\]
using~\cref{eq:nucleon_vector_axial_currents3,eq:neutrino_vector_axial_currents} 
\begin{equation}
\label{eq:scalar_products_helicity}
\begin{aligned}
(l,h^\chi_{++}) & = 8(s-m^2)e^{-i\varphi}\cos^2\frac{\theta}{2}g_R,\\
(l,h^\chi_{--})  & = 8(s-m^2)e^{+i\varphi}\left(g_L-g_R\sin^2\frac{\theta}{2}\frac{m^2}{s}\right),\\
(l,h^\chi_{+-}) & = 8(s-m^2)\frac{m}{\sqrt{s}}\sin\frac{\theta}{2}\cos\frac{\theta}{2}g_R,\\
(l,h^\chi_{-+}) & = -8(s-m^2)\frac{m}{\sqrt{s}}\sin\frac{\theta}{2}\cos\frac{\theta}{2}g_R,
\end{aligned}
\end{equation}
where $g_{L/R}$ are left- and right-handed chirality weak couplings of the nucleon defined in~\cref{eq:left_right_couplings}.

Using~\cref{eq:scalar_products_helicity} and the relationship between Bjorken $y$ and $\sin^2\frac{\theta}{2}$ in the neutrino-nucleon center-of-mass frame
\begin{equation}
\label{eq:Bjorken-y}
y = \frac{(p,q)}{(p,k)}, \quad \sin^2\frac{\theta}{2} = \frac{ys}{s-m^2}
\end{equation}
one can verify that a well known result determining the cross-section of the neutrino-nucleon scattering with $Z^0$-boson exchange is reproduced

\begin{equation}
\begin{aligned}
\sum_{r,r'}&|(l,h^\chi_{r'r})|^2 = 2^6(s-m^2)^2\\
&\times\left(g_L^2+g_R^2(1-y)^2-2g_Lg_R\frac{ym^2}{s-m^2}\right).
\end{aligned}
\end{equation}

\subsection{$\sigma_3$ basis}
We quantize the nucleon's spin along the incoming neutrino direction $\bm{n}_{\bm{k}}$ in the neutrino-nucleon center-of-mass frame. 
This implies that the corresponding two-spinor, which we denote now by $\eta_{r}$, is an eigenvector of the $\bm{n}_{\bm{k}}\times\bm{\sigma}=\sigma_3$ matrix 
\begin{equation}
\sigma_3\eta_r = r\eta_r.
\end{equation}
These two-spinors read
\begin{equation}
\label{eq:two_spinors_fixed_axis}
\begin{aligned}
\eta_{+} & = \begin{pmatrix} 1 \\ 0\end{pmatrix}, 
& \eta_{-} & = \begin{pmatrix} 0\\ 1\end{pmatrix}.\\
\end{aligned}
\end{equation}
Using these two-spinors  instead of $\chi_\pm$ in~\cref{eq:nucleon_vector_axial_currents1,eq:nucleon_vector_axial_currents2}, the nucleon's vector and axial currents, denoting the basis in~\cref{eq:two_spinors_fixed_axis} by the superscript $\eta$, read

\begin{widetext}
\begin{equation}
\label{eq:nucleon_vector_axial_currents4}
\begin{aligned}
V^\eta_{++} & = \left(E_N+m+\cos{\theta}(E_N-m),-\sin{\theta}e^{-i\varphi}E_\nu,-i\sin{\theta}e^{-i\varphi}E_\nu,-(1+\cos{\theta})E_\nu\right),\\
V^\eta_{--} & = \left(E_N+m+\cos{\theta}(E_N-m),-\sin{\theta}e^{+i\varphi}E_\nu,i\sin{\theta}e^{+i\varphi}E_\nu,-(1+\cos{\theta})E_\nu\right),\\
V^\eta_{+-} & = \left(-\sin{\theta}e^{-i\varphi}(E_N-m),(1-\cos{\theta})E_\nu,-i(1-\cos{\theta})E_\nu,\sin{\theta}e^{-i\varphi}E_\nu\right),\\
V^\eta_{-+} & = \left(+\sin{\theta}e^{+i\varphi}(E_N-m),-(1-\cos{\theta})E_\nu,-i(1-\cos{\theta})E_\nu,-\sin{\theta}e^{+i\varphi}E_\nu\right),\\
A^\eta_{++} & =\left(-(1+\cos{\theta})E_\nu,\sin{\theta}e^{-i\varphi}(E_N-m),i\sin{\theta}e^{-i\varphi}(E_N-m),E_N+m+\cos{\theta}(E_N-m)\right),\\
A^\eta_{--}  & = \left((1+\cos{\theta})E_\nu,-\sin{\theta}e^{+i\varphi}(E_N-m),i\sin{\theta}e^{+i\varphi}(E_N-m),-\left(E_N+m+\cos{\theta}(E_N-m)\right)\right),\\
A^\eta_{+-} & = \left(-\sin{\theta}e^{-i\varphi}E_\nu,E_N+m-\cos{\theta}(E_N-m),-i\left(E_N+m-\cos{\theta}(E_N-m)\right),\sin{\theta}e^{-i\varphi}(E_N-m)\right),\\
A^\eta_{-+} & = \left(-\sin{\theta}e^{+i\varphi}E_\nu,E_N+m-\cos{\theta}(E_N-m),i\left(E_N+m-\cos{\theta}(E_N-m)\right),\sin{\theta}e^{+i\varphi}(E_N-m)\right).\\
\end{aligned}
\end{equation}
\end{widetext}
The vector and axial currents calculated in~\cref{eq:nucleon_vector_axial_currents3} and in~\cref{eq:nucleon_vector_axial_currents4} differ because the mathematical and physical sense of $r,r'$ eigenvalues are different.
In~\cref{eq:nucleon_vector_axial_currents3} $r,r'$ are projections of the nucleon spin onto incoming and outgoing momenta of the nucleon, while in~\cref{eq:nucleon_vector_axial_currents4} these are projections onto a fixed axis (chosen to be along the incoming neutrino three-momentum).
One possible illustration is an example of the nucleon scattered backward ($\cos\theta=-1$). 
In this case $(++)$ in the helicity basis represents the nucleon's spin-flip, while in the basis with fixed quantization axis (as defined in~\cref{eq:two_spinors_fixed_axis}), the nucleon's spin does not change its orientation.

Let us briefly review the results obtained in~\cref{eq:nucleon_vector_axial_currents4} to gain further insight. 
For this purpose we  consider three cases most relevant for this paper: (i) the nucleon forward scattering ($\cos\theta=1$), (ii) the nucleon backward scattering ($\cos\theta=-1$) and (iii) the non-relativistic regime ($\sqrt{s}\to m$).

(i) The vector currents 
\begin{equation}
\lim_{\cos\theta\to 1}V^\eta_{++}=\lim_{\cos\theta\to 1}V^\eta_{--}=2(E_N,0,0,-|\bm{P}_N|)
\end{equation}
reduce to the 4-momentum of the nucleon moving towards the incoming neutrino.
The spin-flip in the vector currents are impossible 
\begin{equation}
\lim_{\cos\theta\to 1}V^\eta_{+-}=\lim_{\cos\theta\to 1}V^\eta_{-+}=0.
\end{equation}
On the contrary, the axial current makes the spin-flip possible even for the forward scattering generating non-zero components of the axial current in the transverse plane
\begin{equation}
\begin{aligned}
\lim_{\cos\theta\to 1}A^\eta_{+-}  &= 2m(0,1,-i,0), \\
\lim_{\cos\theta\to 1}A^\eta_{-+}  &= 2m(0,1,i,0).
\end{aligned}
\end{equation}
In this limit there is an exact cancellation of the sum of axial currents with opposite spins
\begin{equation}
\label{eq:axial_sum1}
\lim_{\cos\theta\to 1}\left(A^\eta_{++}+A^\eta_{--}\right)=0.
\end{equation}
This cancellation can be understood by recalling that the axial current of the fermion with the same initial and final momenta $p$ and same spin projection $r$ is proportional to the 4-spin vector $s^\mu$
\begin{equation}
\overline{u}(p,r)\gamma^\mu\gamma^5u(p,r)=2m r s^\mu. 
\end{equation}
Therefore, 
\begin{equation}
\overline{u}(p,+1)\gamma^\mu\gamma^5u(p,+1)+\overline{u}(p,-1)\gamma^\mu\gamma^5u(p,-1)=0,
\end{equation}
which exactly corresponds to~\cref{eq:axial_sum1} for the  forward scattering of the nucleon.
 
(ii) The vector currents conserving spin projection reduce to 
\begin{equation}
\lim_{\cos\theta\to -1}V^\eta_{++}=\lim_{\cos\theta\to -1}V^\eta_{--}=2m(1,0,0,0).
\end{equation}
The spin-flip is also possible 
\begin{equation}
\begin{aligned}
\lim_{\cos\theta\to -1}V^\eta_{+-}&=2|\bm{P}_N|(0,1,-i,0), \\
\lim_{\cos\theta\to -1}V^\eta_{-+}&=2|\bm{P}_N|(0,-1,-i,0)
\end{aligned}
\end{equation}
generating non-zero components of the vector current in the transverse plane.
The axial currents conserving spin projection reduce to  
\begin{equation}
\begin{aligned}
\lim_{\cos\theta\to -1}A^\eta_{++}&=2m(0,0,0,1), \\
\lim_{\cos\theta\to -1}A^\eta_{--}&=2m(0,0,0,-1).
\end{aligned}
\end{equation}
The spin-flip is also possible
\begin{equation}
\begin{aligned}
\lim_{\cos\theta\to -1}A^\eta_{+-}&=2E_N(0,1,-i,0), \\
\lim_{\cos\theta\to -1}A^\eta_{-+}&=2E_N(0,1,i,0).
\end{aligned}
\end{equation}

(iii) The vector current conserves the spin projection
\begin{equation}
\begin{aligned}
\lim_{\sqrt{s}\to m}V^\eta_{++}=\lim_{\sqrt{s}\to m}V^\eta_{--}=2m(1,0,0,0), \\
\lim_{\sqrt{s}\to m}V^\eta_{+-}=V^\eta_{-+}=0.
\end{aligned}
\label{eq:V_non_rel}
\end{equation}
The axial current is non-zero for both cases:  spin projection conserved 
\begin{equation}
\begin{aligned}
\lim_{\sqrt{s}\to m}A^\eta_{++} &= 2m (0,0,0,1), \\
\lim_{\sqrt{s}\to m}A^\eta_{--}  &=2m(0,0,0,-1)
\end{aligned}
\end{equation}
and when the spin-flip occurred
\begin{equation}
\begin{aligned}
\lim_{\sqrt{s}\to m}A^\eta_{+-}&=2m(0,1,-i,0), \\
\lim_{\sqrt{s}\to m}A^\eta_{-+}&=2m(0,1,i,0).
\end{aligned}
\label{eq:A_spin_flip_nonrel}
\end{equation}
One might observe, that in this limit, similar to~\cref{eq:axial_sum1}, 
\begin{equation}
\lim_{\sqrt{s}\to m}\left(A^\eta_{++} + A^\eta_{--}\right) = 0
\end{equation}
which implies that in the coherent term of~\cref{eq:cross-section_3} there is a cancellation of the axial currents for spin-less nuclei.
A more accurate statement can be drawn considering the exact formula in~\cref{eq:nucleon_vector_axial_currents4}
\begin{equation}
\label{eq:axial_sum2}
\begin{aligned}
A^\eta_{++} + A^\eta_{--} &= -2i\sin\theta (E_N-m)(0,\sin\varphi,\cos\varphi,0)\\
                           &\simeq -i\frac{k_0^2}{m}\sin\theta (0,\sin\varphi,\cos\varphi,0)\\
\end{aligned}
\end{equation}
In general, this four-vector is non-zero unless the neutrino energy $k_0$ in the laboratory frame  is not zero and the scattering angle $\theta\ne 0 \text{ or }\pi$.

Once the vector and axial currents of the nucleon are calculated,  it is straightforward to calculate the scalar product $(l,h_{r'r})$ in analogy to the results obtained in the helicity basis~\cref{eq:scalar_products_helicity} 
using~\cref{eq:neutrino_vector_axial_currents,eq:nucleon_vector_axial_currents4}
\begin{equation}
\label{eq:scalar_products_sigma3}
\begin{aligned}
(l,h^\eta_{++}) & = 8(s-m^2)\cos\frac{\theta}{2}\left(g_L-g_R\sin^2\frac{\theta}{2}\frac{m}{\sqrt{s}}(1-\frac{m}{\sqrt{s}})\right),\\
(l,h^\eta_{--})  & = 8(s-m^2)\cos\frac{\theta}{2}\left(1-\sin^2\frac{\theta}{2}(1-\frac{m}{\sqrt{s}})\right)g_R,\\
(l,h^\eta_{+-})  & = -8(s-m^2)e^{-i\varphi}\sin\frac{\theta}{2}\cos^2\frac{\theta}{2}\left(1-\frac{m}{\sqrt{s}}\right)g_R,\\
(l,h^\eta_{-+})  & = 8(s-m^2)e^{i\varphi}\sin\frac{\theta}{2}\\
                            & \times \left(g_L-g_R\frac{m}{\sqrt{s}}\left(1-\sin^2\frac{\theta}{2}(1-\frac{m}{\sqrt{s}})\right)\right),\\
\end{aligned}
\end{equation}
%

%
The scalar products $(l,h^\eta_{rr})$ in~\cref{eq:scalar_products_sigma3} differ from those in~\cref{eq:scalar_products_helicity}.  

(i) The $\varphi$-dependence magically disappeared in the first two lines of~\cref{eq:scalar_products_sigma3} which determine the coherent cross-section.
In $(l,h^\eta_{++})$ it happened because the exponent $e^{i\varphi}$ of the neutrino current in~\cref{eq:neutrino_vector_axial_currents} cancels the exponent $e^{-i\varphi}$ of nucleon currents $V_{++}$ and $A_{++}$ in~\cref{eq:nucleon_vector_axial_currents4}.
In $(l,h^\eta_{++})$ it happened because the corresponding terms, depending now on $e^{i2\varphi}$, cancel each other due to the difference in their relative signs.

The scalars $(l,h^\eta_{+-})$ and $(l,h^\eta_{-+})$ have a $\varphi$-dependence.
However this dependence can not be observed because these terms do not contribute to the coherent cross-section where they could be interfering.

(ii) For the forward scattering $\theta\to 0$, the sum of first two lines in~\cref{eq:scalar_products_sigma3} does not depend on $g_A$
\begin{equation}
\lim_{y\to 0}\left((l,h^\eta_{++}) + (l,h^\eta_{--})\right) = 8(s-m^2) g_V.
\end{equation}
Two other currents with spin-flip vanish in this limit.

(iii) Both scalar products $(l,h^\eta_{rr})$ vanish if the neutrino scatters backward ($\theta=\pi$) in their center-of-mass frame because it corresponds to a change of the total spin of the neutrino-nucleon system by one unit.
This result explains why the coherent cross-section vanishes when the recoil nucleus has the maximum kinetic energy.

(iv) In the non-relativistic limit
\begin{equation}
\label{eq:scalar_products_sigma4}
\begin{aligned}
\lim\limits_{\sqrt{s}\to m}\frac{(l,h^\eta_{++})}{8(s-m^2)} & = \cos\frac{\theta}{2}g_L,\\
\lim\limits_{\sqrt{s}\to m}\frac{(l,h^\eta_{--})}{8(s-m^2)}  & = \cos\frac{\theta}{2}g_R,\\
\lim\limits_{\sqrt{s}\to m}\frac{(l,h^\eta_{+-})}{8(s-m^2)}  & = 0,\\
\lim\limits_{\sqrt{s}\to m}\frac{(l,h^\eta_{-+})}{8(s-m^2)}  & = e^{i\varphi}\sin\frac{\theta}{2}g_A,\\
\end{aligned}
\end{equation}
where we kept the $8(s-m^2)$ factor in the denominators  because it is precisely canceled in the calculation of the corresponding cross-section.

It is interesting to observe two distinct limits of the last two lines in~\cref{eq:scalar_products_sigma4} for similar spin-flipped configurations. 
In the non-relativistic limit the vector currents with spin-flip vanish as can be seen in~\cref{eq:V_non_rel} and only the transverse components of the axial currents in~\cref{eq:A_spin_flip_nonrel} survive.
However, the axial current $A^\eta_{+-}$ in~\cref{eq:A_spin_flip_nonrel} turned out to be orthogonal to the neutrino current at $\sqrt{s}\to m$.
At the same time, the scalar product of the latter and of $A^\eta_{-+}$ from~\cref{eq:A_spin_flip_nonrel} is non-zero and proportional to the axial coupling $g_A$.

\subsection{Relation between helicity and $\sigma_3$ bases}
The eigenvectors $\chi_\pm$ and $\eta_\pm$ in the helicity and $\sigma_3$ bases, defined in~\cref{eq:helicity_basis,eq:two_spinors_fixed_axis}, respectively, are related to each other via a linear transformation
\begin{equation}
\label{eq:relation_bases}
\begin{aligned}
\begin{pmatrix}
\eta_{+}\\
\eta_{-} 
\end{pmatrix}
&=
\begin{pmatrix}
0 & 1\\
-1 & 0
\end{pmatrix}
\begin{pmatrix}
\chi_{+}(\bm{n}_{\bm{p}})\\
\chi_{-}(\bm{n}_{\bm{p}})
\end{pmatrix}\\
&=
\begin{pmatrix}
\sin\frac{\theta}{2}                            &\cos\frac{\theta}{2}e^{i\varphi}\\
-\cos\frac{\theta}{2}e^{-i\varphi}     & \sin\frac{\theta}{2}  
\end{pmatrix}
\begin{pmatrix}
\chi_{+}(\bm{n}_{\bm{p}'})\\
\chi_{-}(\bm{n}_{\bm{p}'})
\end{pmatrix}.
\end{aligned}
\end{equation}
\cref{eq:relation_bases} allows one to relate the nucleon currents and scalar products $(l,h)$ calculated in the two bases
\begin{equation}
\label{eq:relation_bases_plus}
\begin{aligned}
(l,h^\eta_{++}) & = \sin\frac{\theta}{2}(l,h^\chi_{+-})+\cos\frac{\theta}{2}e^{-i\varphi}(l,h^\chi_{--}),\\
(l,h^\eta_{-+}) & = -\cos\frac{\theta}{2}e^{i\varphi}(l,h^\chi_{+-})+\sin\frac{\theta}{2}(l,h^\chi_{--})\\
\end{aligned}
\end{equation}
and
\begin{equation}
\label{eq:relation_bases_minus}
\begin{aligned}
(l,h^\eta_{+-}) & = -\sin\frac{\theta}{2}(l,h^\chi_{++})-\cos\frac{\theta}{2}e^{-i\varphi}(l,h^\chi_{-+}),\\
(l,h^\eta_{--}) & = \cos\frac{\theta}{2}e^{i\varphi}(l,h^\chi_{++})-\sin\frac{\theta}{2}(l,h^\chi_{-+}).\\
\end{aligned}
\end{equation}

(i) The nucleon currents, for example, in the left-hand-side of the first line in~\cref{eq:relation_bases_plus}  correspond to the scatterings when the spin projection on the fixed axis (incoming neutrino direction) does not change ($h^\eta_{++}$) and is flipped  ($h^\eta_{-+}$). 
These currents can be uniquely described by linear combinations of the current with negative helicity in both initial and final states $h^\chi_{--}$ and of the current in which the initially negative helicity is flipped $h^\chi_{+-}$, as illustrated in Fig.~\ref{fig:spin_decomposition}. 
\begin{figure}[!h]
	\resizebox{\linewidth}{!}{\input{spin.tex}}
	\caption{
The left pictogram corresponds to a fermion current $h^\eta_{++}$ with positive projections of its spin (shown by double arrowed line) on the given axis (shown by dashed line) in both initial and final states.
This current is decomposed into a sum of the current with negative helicity in both the initial and final states $h^\chi_{--}$ and of the current in which the initially negative helicity is flipped $h^\chi_{+-}$ weighted with $\sin\frac{\theta}{2}$ and $\cos\frac{\theta}{2}e^{-i\varphi}$, respectively, as illustrated by the right pictogram. 
}
\label{fig:spin_decomposition}
\end{figure}

(ii) The following equalities hold true
\begin{equation}
\label{eq:relation_bases_squares}
\begin{aligned}
|(l,h^\eta_{++})|^2+|(l,h^\eta_{-+})|^2 & = |(l,h^\chi_{+-})|^2+|(l,h^\chi_{--})|^2,\\
|(l,h^\eta_{+-})|^2+|(l,h^\eta_{--})|^2 & = |(l,h^\chi_{++})|^2+|(l,h^\chi_{-+})|^2.
\end{aligned}
\end{equation}

(iii) \cref{eq:relation_bases_plus,eq:relation_bases_minus} can also be used to cross-check the results of tedious calculations leading to~\cref{eq:nucleon_vector_axial_currents4}.

\section{An analogy in mechanical system for elastic and inelastic scattering}
\label{app:mechanics}
Kinematic issues discussed in~\cref{sec:kinematic_paradox} are valid not only for a quantum system but also for a mechanical system.
As a useful illustration of  elastic and inelastic scattering let us consider a system of two balls with equal masses $m$ connected by a massless spring with non-zero rigidity.

(i) Consider this system when both balls  are at rest and one ball gains a momentum $\bm{q}$, and thus the kinetic energy $\bm{q}^2/2m$.
After the acquired momentum is redistributed among the balls, their center-of-mass  moves with momentum $\bm{q}$ but with a kinetic energy two time smaller, $\bm{q}^2/4m$, because of its mass $2m$.

\begin{figure}[!h]
	\resizebox{\linewidth}{!}{\tikzset{
  particlepath/.style = {decoration={markings,mark=at position 0.50 with {\arrow[xshift=0.8mm]{stealth}}},postaction={decorate}},
  momentum/.style     = {-stealth},
  boson/.style={decorate,decoration={snake}},
  nuclint/.style={decorate,decoration={snake,amplitude=0.5mm,segment length=2mm}},
  secondary/.style={opacity=0.15}
}
{\scalefont{3.5}
\begin{tikzpicture}[very thick]
    \def\Radius{20mm}
    \def\Length{40mm}
    \def\Amplitude{4mm}
    \def\Segment{12mm}
    \def\Skip{25mm}
    \def\VecLength{10mm}

    \def\Plot#1#2#3#4#5{%
    \tikzset{
      spring/.style={decorate,decoration={snake,amplitude=\Amplitude/#2,segment length=#2*\Segment}},
      label/.style={above,yshift=\Radius}
      }
      \draw #1 circle [radius=\Radius] node[label] {#3} ++(0:\Radius) coordinate (left);
      \draw[spring] (left) -- ++(0:#2*\Length) coordinate(right) node[label,midway] {#5};
      \draw (right) ++(0:\Radius) circle [radius=\Radius] node[label] {#4} coordinate (end);
    }

    \Plot{(0,0)}{1}{$\bm{p}=\bm{0}$}{$\bm{p}=\bm{0}$}{$\bm{p}_\text{cm}=0$}
    \draw[-stealth] (end) ++(180:\VecLength) -- ++(0:\VecLength) node [midway,below] {$+\bm{q}$};
    \draw[-{Stealth[length=4mm]},ultra thick] (end) ++(0:\Skip) -- ++(0:\Skip) coordinate (end);
    \Plot{(16,+2)}{0.5}{}{}{$\bm{p}_\text{cm}=\bm{q}$}
    \Plot{(14,-3)}{1.5}{}{}{}

  \end{tikzpicture}
}}
	\caption{Both balls are initially at rest when the right ball is hit with a momentum $\bm{q}$ (left).
	After some time their center-of-mass  moves with momentum $\bm{q}$ and half of the transferred energy is accumulated in the potential energy of the spring, shown by its tension and extension (right).
	 }
	\label{fig:mech_inc}
\end{figure}
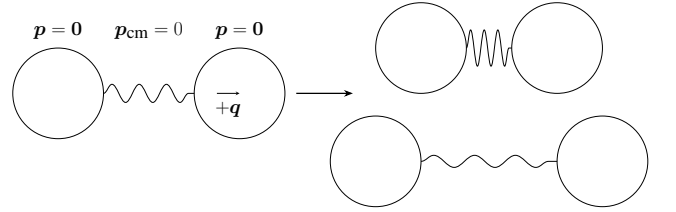

Half of the initial kinetic energy went into the potential energy of the spring.
In this analogy the initial ground state is transformed into an excited state. 
This example, shown in~\cref{fig:mech_inc}, illustrates an inelastic scattering.

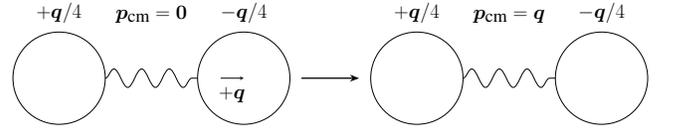
\begin{figure}[!h]
	\resizebox{\linewidth}{!}{\tikzset{
  particlepath/.style = {decoration={markings,mark=at position 0.50 with {\arrow[xshift=0.8mm]{stealth}}},postaction={decorate}},
  momentum/.style     = {-stealth},
  boson/.style={decorate,decoration={snake}},
  nuclint/.style={decorate,decoration={snake,amplitude=0.5mm,segment length=2mm}},
  secondary/.style={opacity=0.15}
}
{\scalefont{3.5}
\begin{tikzpicture}[very thick]
    \def\Radius{20mm}
	\def\Length{40mm}
	\def\Amplitude{4mm}
	\def\Segment{12mm}
	\def\Skip{25mm}
	\def\VecLength{10mm}

    \def\Plot#1#2#3#4#5{%
    \tikzset{
      spring/.style={decorate,decoration={snake,amplitude=\Amplitude/#2,segment length=#2*\Segment}},
      label/.style={above,yshift=\Radius}
      }
      \draw #1 circle [radius=\Radius] node[label] {#3} ++(0:\Radius) coordinate (left);
      \draw[spring] (left) -- ++(0:#2*\Length) coordinate(right) node[label,midway] {#5};
      \draw (right) ++(0:\Radius) circle [radius=\Radius] node[label] {#4} coordinate (end);
    }

    \Plot{(0,-40mm)}{1}{$+\bm{q}/4$}{$-\bm{q}/4$}{$\bm{p}_\text{cm}=\bm{0}$}
    \draw[-stealth] (end) ++(180:\VecLength) -- ++(0:\VecLength) node [midway,below] {$+\bm{q}$};
    \draw[-{Stealth[length=4mm]},ultra thick] (end) ++(0:\Skip) -- ++(0:\Skip) coordinate (end);
    \Plot{(end) ++(0:\Skip)}{1}{$+\bm{q}/4$}{$-\bm{q}/4$}{$\bm{p}_\text{cm}=\bm{q}$}

  \end{tikzpicture}
}}
	\caption{Balls move towards and away each other with maximal momenta $\pm\bm{q}/4$ while their center-of-mass is at rest, when the right ball having  momentum $-\bm{q}/4$ is hit and acquired the momentum $\bm{q}$ (left).
		After a while, their center-of-mass  moves with momentum $\bm{q}$, while the spring still has the same potential energy, remaining in the same state.
		The balls have the same maximal momenta $\pm\bm{q}/4$.
	}
	\label{fig:mech_coh}
\end{figure}

(ii) Consider now the same balls in an initially excited state, i.e. moving towards and away from each other while their center-of-mass is at rest.
The momenta of these balls at any time are equal to each other in magnitude, and have opposite directions.
Let the maximum momentum of a ball, reached when the spring has zero potential energy, be equal to $\pm\bm{q}/4$  as shown in~\cref{fig:mech_coh}. 
One of these momenta is chosen  in agreement with~\cref{eq:nucleon_momentum_coherent} to be $-\bm{q}/4$.
The total energy of both balls equals $(\bm{q}/4)^2/2m+(\bm{q}/4)^2/2m=\bm{q}^2/16m$. 
Let the ball with a momentum $-\bm{q}/4$ acquire an additional momentum $\bm{q}$.

Right after this interaction, the untouched and hit balls have  momenta $(+\bm{q}/4,+3/4\bm{q})$  and energies $(\bm{q}^2/32m,9\bm{q}^2/32m)$, respectively.
The total accumulated energy is equal to $5\bm{q}^2/16m$.

Since the kinetic energy of their center-of-mass equals $\bm{q}^2/4m$, one finds that the potential energy accumulated in the spring  is  unchanged $\bm{q}^2/16m$.

In this example, the system of two balls connected by a spring remains in the state with the same internal potential energy.
The change of kinetic energy of the struck ball  is exactly the kinetic energy  of their center-of-mass.
This is an example of elastic scattering.
\bibliographystyle{apsrev4-1}
\bibliography{COHERENT}

\end{document}